\documentclass{article}

\usepackage{microtype}
\usepackage{graphicx}
\usepackage{subfigure}
\usepackage{amsmath}
\usepackage{booktabs} 
\usepackage{amssymb}
\usepackage{hyperref}
\usepackage{url}
\usepackage{threeparttable, multirow}
\usepackage{listings}

\usepackage[accepted]{icml2024}

\usepackage{xspace}
\newcommand{\prjname}{KV-Runahead\xspace}

\icmltitlerunning{\prjname: Scalable Causal LLM Inference  by Parallel Key-Value Cache Generation}

\begin{document}

\twocolumn[
\icmltitle{\prjname: 
Scalable Causal LLM Inference  by\\Parallel Key-Value Cache Generation}




\icmlsetsymbol{equal}{*}

\begin{icmlauthorlist}
\icmlauthor{Minsik Cho}{apple}
\icmlauthor{Mohammad Rastegari}{meta}
\icmlauthor{Devang Naik}{apple}

\end{icmlauthorlist}

\icmlaffiliation{apple}{Apple. USA} 
\icmlaffiliation{meta}{Meta. USA (the work done while being with Apple)} 
\icmlcorrespondingauthor{Minsik Cho}{minsik@apple.com}
\icmlkeywords{Machine Learning, ICML}

\vskip 0.3in
]



\printAffiliationsAndNotice{}  

\begin{abstract}
Large Language Model  or LLM  inference has two phases, 
the prompt (or prefill) phase to output the first token and the extension (or decoding) phase to the generate subsequent tokens.
In this work, we propose an efficient parallelization scheme, \prjname to accelerate the prompt phase.
The key observation is that the extension phase generates tokens faster than the prompt phase because of  key-value cache (KV-cache).
Hence, 
\prjname parallelizes the prompt phase by orchestrating multiple processes to populate the KV-cache and minimizes the   time-to-first-token  (TTFT).
Dual-purposing the KV-cache scheme has two  main benefits. 
First, since KV-cache is designed to leverage the causal attention map, we  minimize computation and computation automatically.  Second, since
it already exists for the extension phase, \prjname is easy to implement.
We further propose context-level load-balancing  to handle uneven  KV-cache generation  (due to the causal attention) and to optimize TTFT.
Compared with an existing parallelization scheme such as    tensor or sequential parallelization where keys and values are locally generated and exchanged via all-gather collectives, 
our experimental results demonstrate that \prjname can offer over 1.4$\times$ and 1.6$\times$ speedups for Llama 7B and Falcon 7B respectively.

\end{abstract}
\section{Introduction}
Large language models or LLMs, and especially Generative Pre-trained Transformer  (GPT) models have shown excellent performance on many complex language tasks~\cite{ouyang2022training, zhang2022opt}. 
However, the decoder architecture and autoregression execution in LLMs
pose two challenges for efficient inferences:
\textbf{a) Time-to-first-token or TTFT}: consuming potentially a long user context and generate the first token  \textbf{b) Time Per Output Token or TPOT}: generating the subsequent tokens fast~\cite{liu2023cachegen}.
The second challenge is known to be a memory-bound problem, and a large body of research has been done~\cite{pope2022efficiently}, including sparsification, quantization, or weight clustering~\cite{frantar2023gptq,lin2023awq,cho2023edkm,liu2023llmqat} or speculative decoding~\cite{leviathan2023fast,chen2023accelerating}. But, the first challenge for a long user context is largely a compute-bound problem~\cite{liu2023cachegen,NVidia-LLM-Inference} and 
 critical for favorable user experience with retrieval augmentation~\cite{ram2023incontext},
in-context learning~\cite{dong2023survey}, summarization~\cite{zhang2023benchmarking}, story generation~\cite{zhang2023survey}, and so on.

\begin{figure*}[!t]
\centering
   \centering
\mbox{
\hspace{-0.1 in}
		\subfigure[LLM inference consists of two phases: the prompt phase for the first token and the extension phase for the next tokens.]
		{\includegraphics[width=3.4 in]{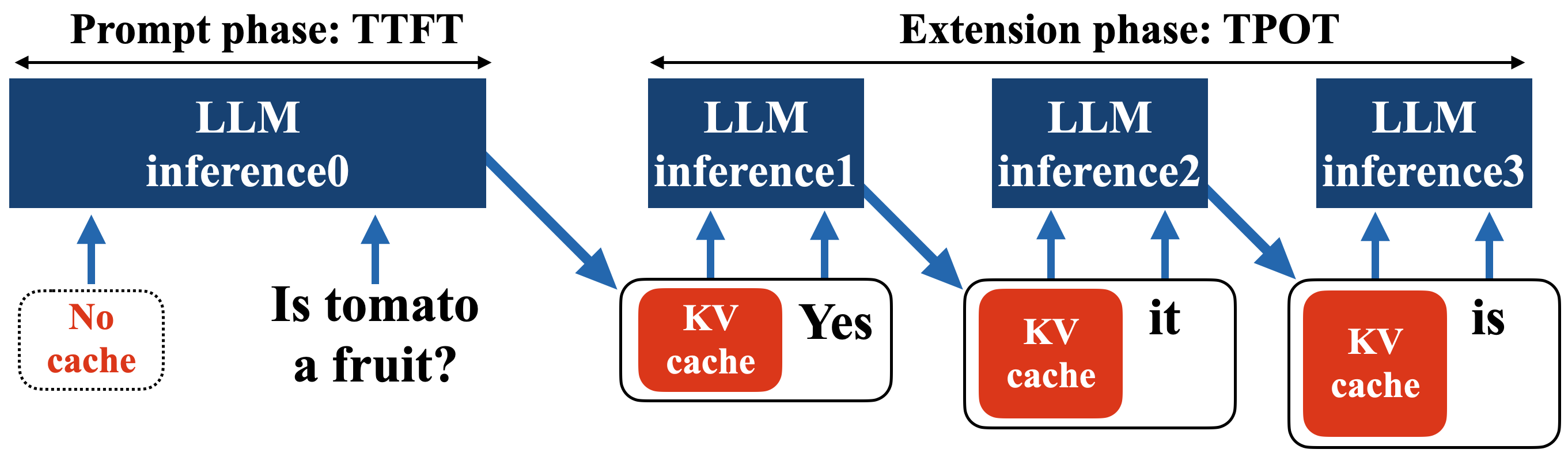}} 
 
  \hspace{0.1 in}
		\subfigure[Causal attention computation of one layer on one process: the upper triangle of $QK^T$ will be masked out with $-\infty$.]
		{\includegraphics[width=3.0 in]{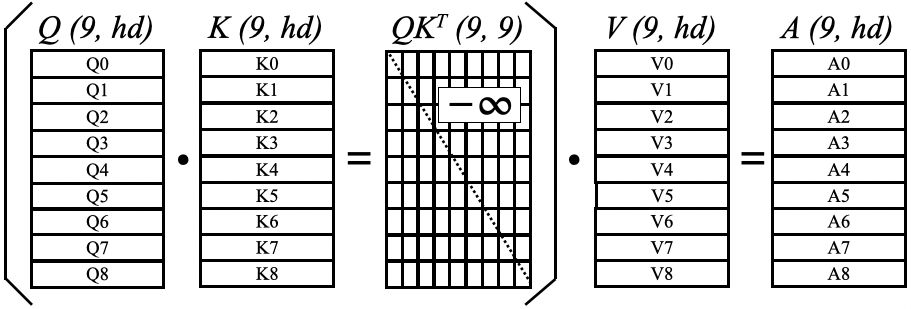}} 
 }  
 
\vspace{-0.10 in}
\caption{ 
LLM inference begins with the prompt phase to generate the KV-cache and the first token which drive  the extension phase as in (a). Inside each layer of the LLM as in (b), a causal attention map ($QK^T$) is built to compute the attention $A$ from query, value, and key s $(Q,K,V)$. Computing attention thus has $O(C^2)$ complexity where $C$ is the user context.}
\label{llm_infer}    
\vspace{-0.10 in}
\end{figure*}

Since TTFT for a long context is compute-bound, one solution is to use more computing power in the form of parallelization. The current SOTA in LLM parallelization inlcudes tensor and sequential parallelization~\cite{patel2023splitwise, li-etal-2023-sequence,korthikanti2022reducing,NVidia-LLM-Inference} where the key and value computations are distributed over multiple processes and subsequently exchanged, 
aiming to compute the attention map perfectly in parallel. The methods above are generic enough to drive  LLM inference~\cite{attention}, 
but not specialized enough for scalable LLM inference, as the causality in attention is
not fully leveraged, resulting in up to 2$\times$ overhead in terms of both computation and communication over the ideal case.



Therefore, we propose a novel yet effective parallelization technique tailed for LLM inference, \prjname to minimize TTFT.
By re-purposing the key-value cache or KV-cache~\cite{NVidia-LLM-Inference} mechanism (which exists anyway for subsequent token generation), our proposed \prjname uses other processes to populate KV-cache for the last process with context-level load-balancing. Since KV-cache assumes causal attention computation, \prjname reduces the computation and communication costs and offers lower TTFT over the existing methods. Further, \prjname  requires minimal engineering costs, as it simply makes the KV-cache interface dual-purposed. In detail, our contributions are the following:
 
\begin{itemize}  
\vspace{-0.1 in}
   \item \textbf{We demonstrate that  KV-cache scheme can be dual-purposed to parallelize LLM inference for low TTFT}. Since the KV-cache is built on the causal decoder and gets populated in parallel, \prjname can offer considerable compute and communication savings over tensor/sequential parallelization.
   \item \textbf{We show that using KV-cache for parallelization enables asynchronous communication}. Thus,   
   \prjname replaces  global synchronization with point-to-point asynchronous communication, and provides robustness against  network bandwidth fluctuation.
   \item \textbf{We highlight that context-level partitioning  can load-balance parallel LLM inference.} Asymmetric computations and communication rise from  KV-cache and its dependency chain across parallel processes. Yet, we can mitigate the negative effects on TTFT with the proposed context-level load-balancing.

   \item \textbf{We propose that hierarchical grid search 
    for efficient context-partitioning}. Such search results contribute to a lookup table from which a TTFT-minimizing partitioning can be interpolated for various context lengths.
   
 
\end{itemize}

\section{Related Works}
\label{related}
\textbf{LLM Inference:} Generative LLM inference consists of two steps as in Fig.~\ref{llm_infer}~\cite{patel2023splitwise}. Once the user context is received, all the input tokens are consumed  to generate the first token, which is called   the prompt phase. At the same time, the computed key and value embeddings are saved as  KV-cache~\cite{optimus,liu2023cachegen} and fed to all subsequent token generations to expedite the extension phase. Accordingly, KV-cache grows as more tokens are generated, because the next token generation needs to attend to all previous tokens, including the user context.
While the critical metric for the extension phase is time-per-output-token or TPOT, the prompt phase needs to deliver the first token fast which is measured as time-to-first-token or TTFT.

\textbf{TTFT Optimization:}
Minimizing TTFT, especially for long context requires two efforts: efficient KV-cache management and fast attention map computation.
PagedAttention~\cite{kwon2023efficient} facilitates the exchange of data including KV-cache   between different memory subsystems to handle long contexts.
Infinite-LLM~\cite{lin2024infinitellm} suggests distributed KV-cache management system at the cloud scale to adaptively handle extremely long context lengths.
CacheGen~\cite{liu2023cachegen} proposes compressing KV-cache for pre-computed contexts to lower TTFT.
SplitWise~\cite{patel2023splitwise} proposes to use two different platforms, one with high computing capacity for the prompt phase and the other with low computing capacity for the extension phase by transferring the LLM states, including KV-cache from the first to the second platforms.

\begin{figure}[!b]
\centering
   \centering
\mbox{
\hspace{-0.1in}
		\subfigure[1 process]
		{\includegraphics[width=0.8 in]{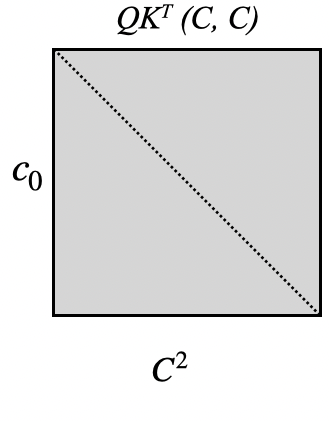}} 
		\subfigure[2 processes]
		{\includegraphics[width=0.8 in]{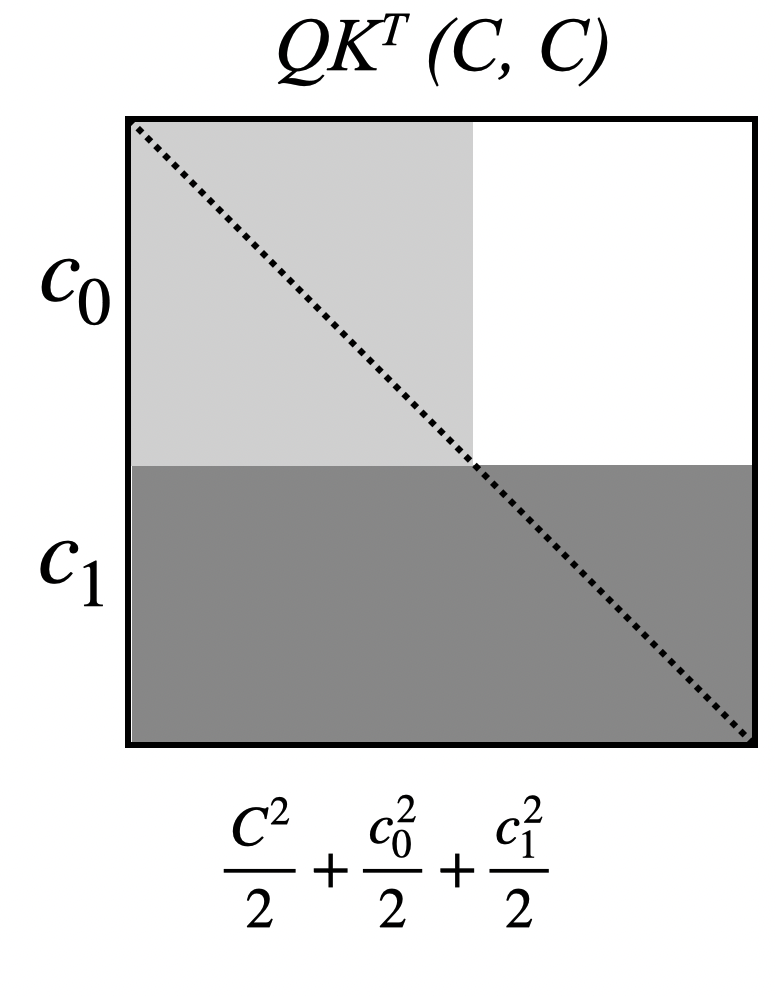}}  

		\subfigure[4 processes]
		{\includegraphics[width=0.8 in]{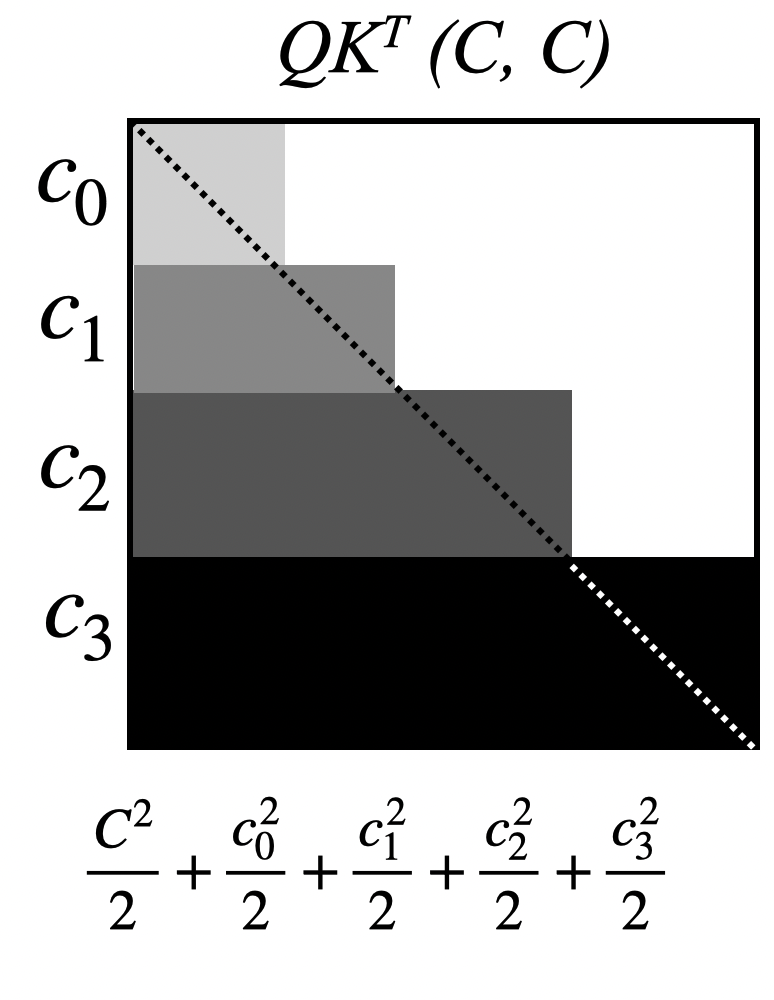}} 

		\subfigure[$p$ processes]
		{\includegraphics[width=0.8 in]{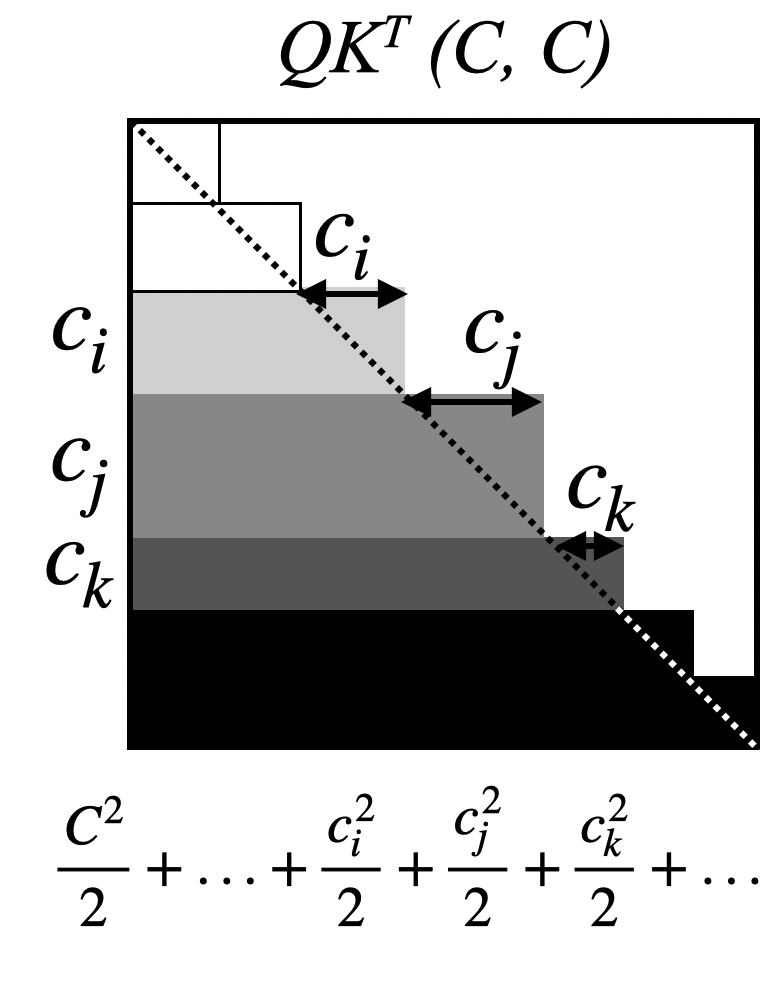}}   
 }  
  \vspace{-0.15 in}
\caption{ 
$QK^T$ computation coverage using BLAS matrix-matrix multiplication: 
by linking each context partition to KV-cache, 
we can closely approximate the lower triangular part and minimize unnecessary dot products. Note the upper triangular part of the attention will be masked out to enforce causality.}
\label{kvr:causal_attn}    
\end{figure}

\begin{figure*}[!t]
\centering
   \centering
\mbox{

		\subfigure[Tensor+Sequence Parallel Inference: With even context partitioning, each process will perform \textit{all-gather} over  $K,V$.   All computations in the layers are symmetric and globally synchronized, and the user context can be evenly partitioned.   (See Fig.~\ref{dp::before_after}).]
		{\includegraphics[width=3.2 in]{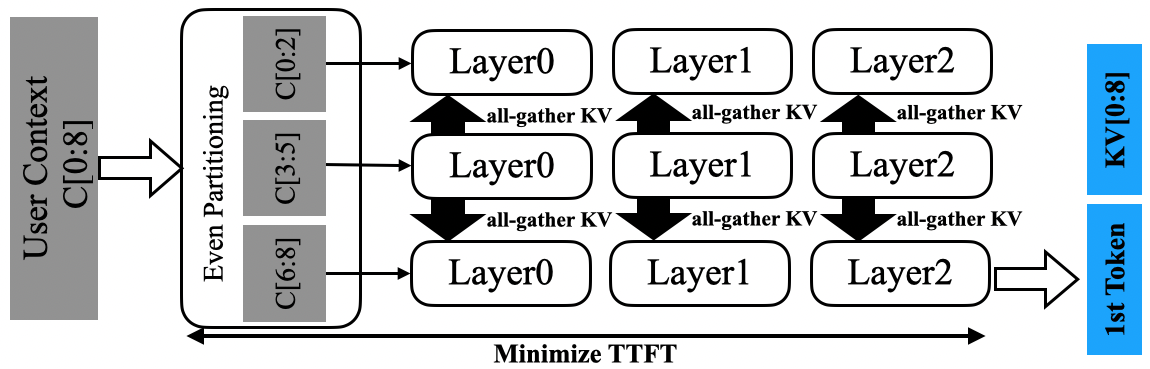}} 
 \hspace{0.2 in}
		\subfigure[\prjname Inference: Since the later processes wait for the KV-cache to be ready, the layers will asynchronously communicate, and the user context is unevenly partitioned (for load-balancing) to minimize TTFT.  (See Fig.~\ref{kvr::before_after}).]
		{\includegraphics[width=3.2 in]{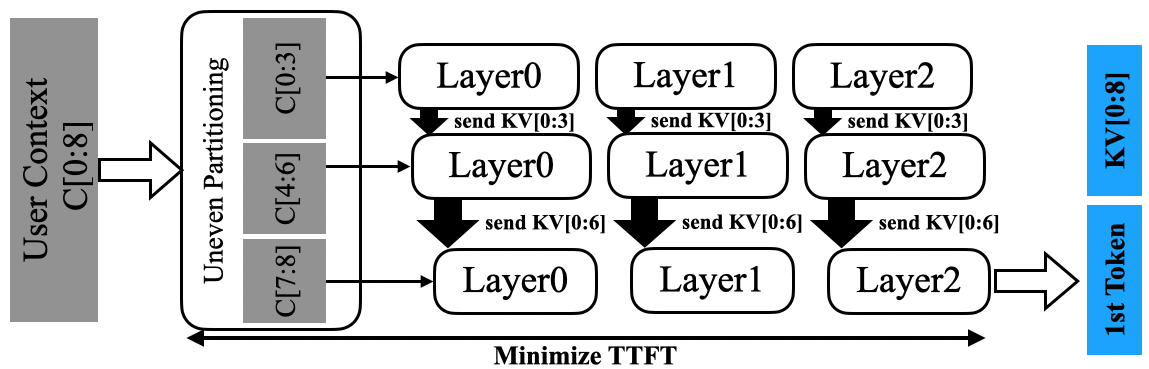}} 
 }   

\vspace{-0.15 in}
\caption{ 
Comparing the existing tensor+sequence parallel scheme with the proposed \prjname for parallel LLM inference.}
\label{kvr:overview}
\vspace{-0.15 in}
\end{figure*}

\textbf{LLM Inference Parallelization:} Since TTFT optimization is compute-bound, one can employ parallel DNN inference.
Pipeline parallelism shards the layers of a model across  multiple processes, splitting the model into several stages or layers~\cite{huang2019gpipe,pmlr-v139-narayanan21a, agrawal2023sarathi}.
Tensor Parallelism is one of the popular parallel methods from~\cite{HuggingFace-Tensor-Parallelism,shoeybi2020megatronlm,narayanan2021efficient} where a large matrix multiplication is scattered and then the partial output matrices are gathered, and is known to be superior to pipeline parallelism~\cite{patel2023splitwise}.
Sequence parallelization~\cite{NVidia-LLM-Inference, li-etal-2023-sequence} is a novel data parallel algorithm (by evenly partitioning the input sequences over multiple processes) coupled with a distributed ring attention algorithm. 
By deploying the ring topology over all the devices,  each process exchanges the key and value embedding with neighbors and builds a full attention map locally. 

Both tensor and sequence parallelizations in LLM are mathematically similar in a sense that \textbf{a)} one of two matrices (i.e., either activations or parameters) in  multiplication will be sharded over multiple devices, \textbf{b)} both require collective communication to merge the partial outcomes. Hence, both
are popular for parallel LLM inference~\cite{korthikanti2022reducing}, yet not specialized enough for causal attention, leading to excessive   computation and communication overheads.




\section{Causal LLM Scalability and Motivation}
\label{motiv}

In this section, we will discuss the lower bound of the scalability of a causal attention-based LLM for a sufficiently long user context $C$ over parallel  $p$ processes. 
Assume that the user context $C$ is partitioned into $C = \{c_0, c_1, c_2, ..., c_{p-1}\}$ for $p$ processes, and each process is exclusively mapped to one compute fabric (e.g., GPU). The minimum compute over $p$ to generate the first token, $TTFT(p)$   with perfect load-balancing is as follows:
\begin{align}
TTFT(p) &\ge \alpha \left[ \frac{\frac{1}{2}C^2 + \frac{1}{2}(\sum^{p-1}_0 c_i^2)}{p} \right]   \label{kvr:attn_p_s0} \\
&\ge  \alpha \left[ \frac{\frac{1}{2}C^2 +   \frac{1}{2}p(\frac{C}{p})^2}{p} \right] \nonumber \\
& = \alpha \left[ \frac{C^2}{2} (\frac{1}{p} + \frac{1}{p^2}) \right] \nonumber \\
& =  \frac{TTFT(1)}{2}(\frac{1}{p} + \frac{1}{p^2})\\
& = \textbf{TTFT$^{*}$}(p)
\label{kvr:attn_p}
\end{align}
where $\alpha$ is a fitting coefficient such that $TTFT(1) = \alpha C^2$ (single process performance)~\cite{dao2022flashattention}, and \textbf{TTFT$^{*}$}(p) is the lower bound of TTFT over $p$.
The significance of $\textbf{TTFT$^{*}$}(p)$ is that  for a very long user context, there exists super-linear scalability (i.e., more than 2$\times$ speedup with 2 processes)   with causal LLM in the ideal setup, such as perfect load-balancing, zero communication costs, and so on. Please see the super-linear scalability of \prjname reported in Fig.~\ref{ttft:result_llama7b} (d).

Fig.~\ref{kvr:causal_attn} visualizes the concepts behind Eq.~(\ref{kvr:attn_p_s0}) which essentially divides  an attention map, $QK^T(C,C)$ in the shaded regions over $p$ processes. 
We need to practically compute multiple rectangle regions using matrix-matrix multiplication and mask out the upper triangle part (which is how most   LLMs are implemented). Therefore, with more partitions, we can eliminate the wasted computation.
Other equally good partitioning setups (i.e., using vertical rectangles to approximate the lower triangle) could exist, but the one in Fig.~\ref{kvr:causal_attn} (d) is LLM-friendly: easy to generate at the context level, and exactly aligned with KV-cache.

Hence, we can intuitively map the partitions in Fig.~\ref{kvr:causal_attn} (d) to $p$ processes, which can be   implemented by dual-purposing the already-existing KV-cache interface with minor efforts, leading to the motivation behind ~\prjname. Also, as seen in Fig.~\ref{kvr:causal_attn} (d),  each process will suffer from varying computation load, thus one may not effectively minimize TTFT. Yet, optimizing $c_i$ alone may lead to global over-computation.
Hence, we perform context-level partitioning for load-balancing and minimal TTFT in \prjname.





\begin{figure*}[!t]
\centering
   \centering
\mbox{

		\subfigure[$Q, K, V$ are computed for the evenly partitioned context.]
		{\includegraphics[width=3.0 in]{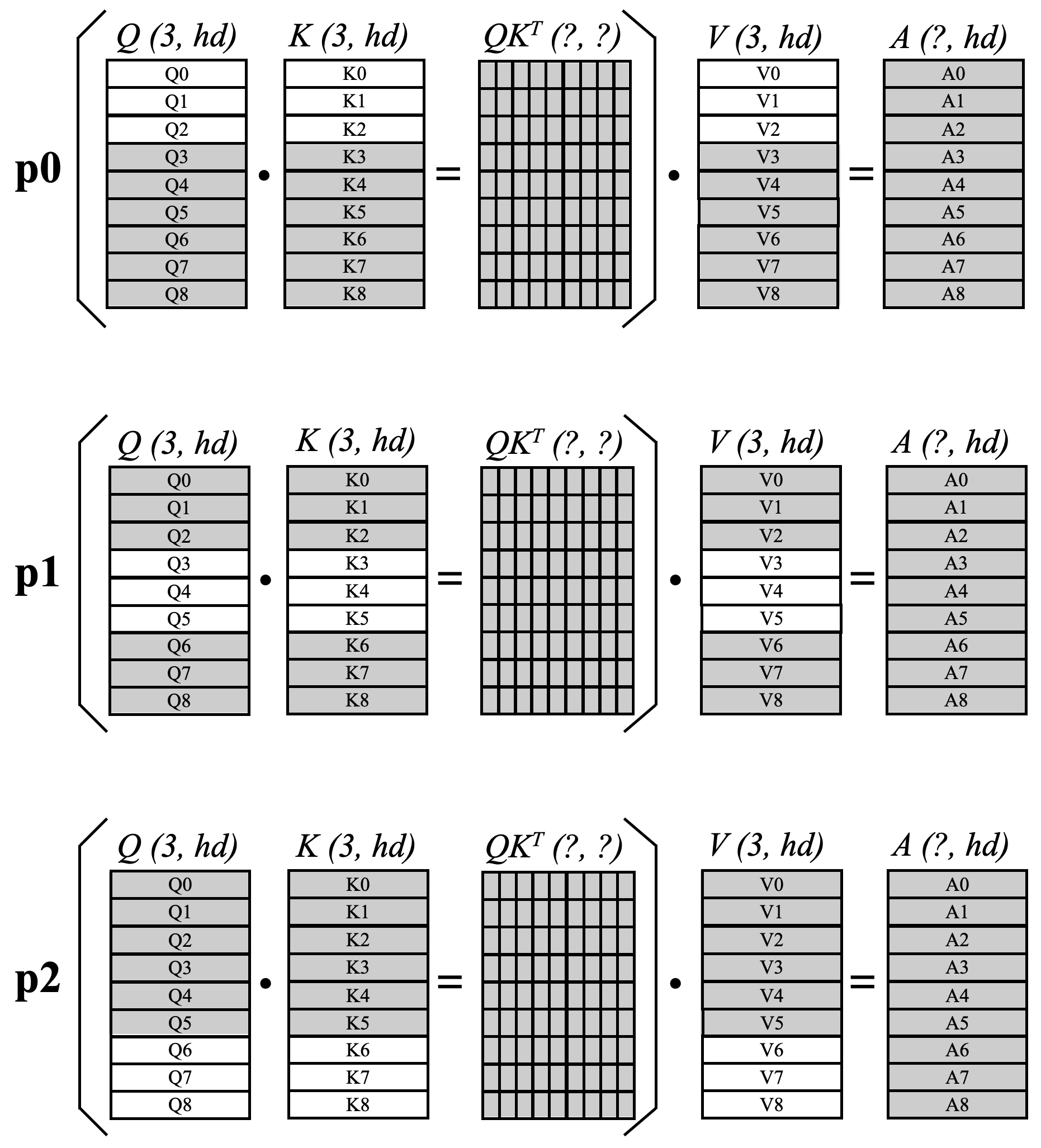}} 
 
  \hspace{0.6 cm}
		\subfigure[$K, V$ are exchanged by \textit{all-gather} to mimic Fig.~\ref{llm_infer} (b). ]
		{\includegraphics[width=2.85 in]{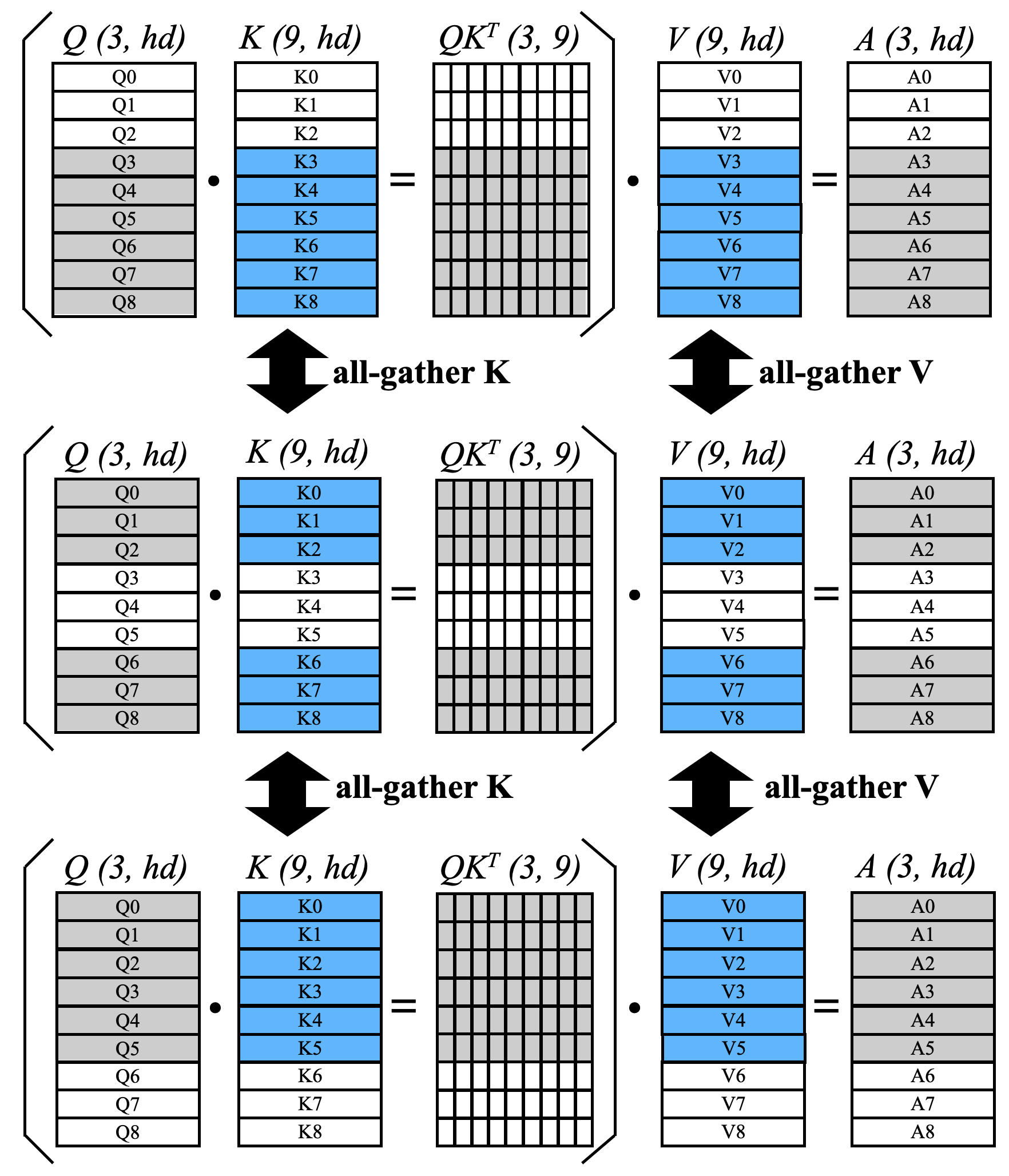}} 
 }  
 
\vspace{-0.15 in}
\caption{ 
Tensor/sequence parallel inference over 3 processes $p_{\{0,1,2\}}$ within a layer to compute attention map ($QK^T$) and final attention $A$: Each process will compute the equal amount of $(Q,K,V)$ in (a), and then
globally share ($K,V$) using \textit{all-gather} collectives to compute the equally sized partial $QK^T$ (i.e.,   \textbf{27 dot-products} needed on each) and partial $A$. Such \textit{all-gather} operations require  global synchronization, and incur the  \textbf{traffic} for \textbf{36} $(K,V)$ entries (i.e., the  number of blue rows in $(K,V)$).}
\label{dp::before_after}    
\end{figure*}

\begin{figure*}[!t]
\centering
   \centering 

\mbox{

		\subfigure[$Q, K, V$ are computed for the uneven partitioning.]
		{\includegraphics[width=3.0 in]{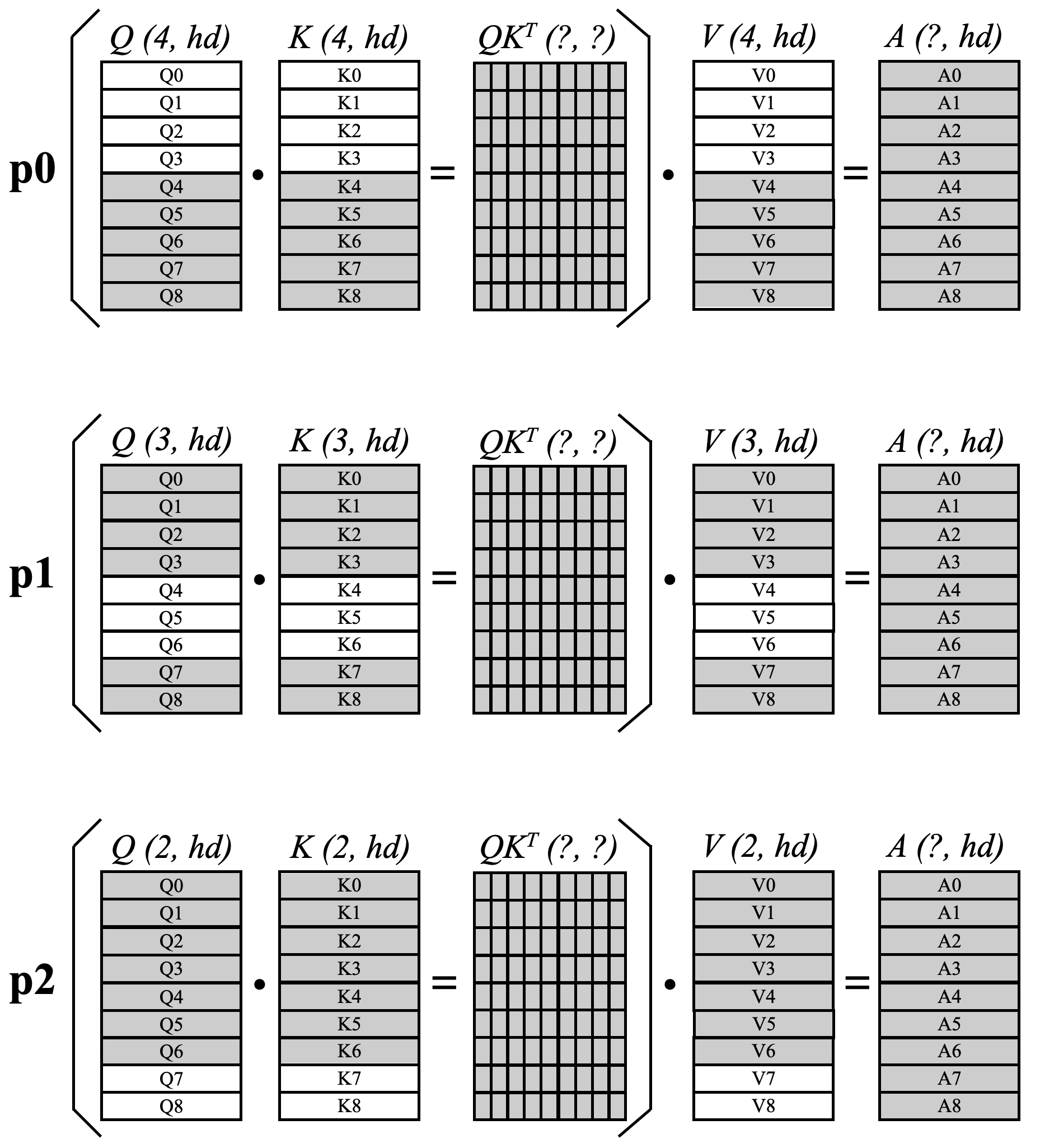}} 
 
  \hspace{0.6 cm}
		\subfigure[KV-caches are handed down from $p_i$ to $p_{i+1}$ by \textit{send}. ]
		{\includegraphics[width=2.85 in]{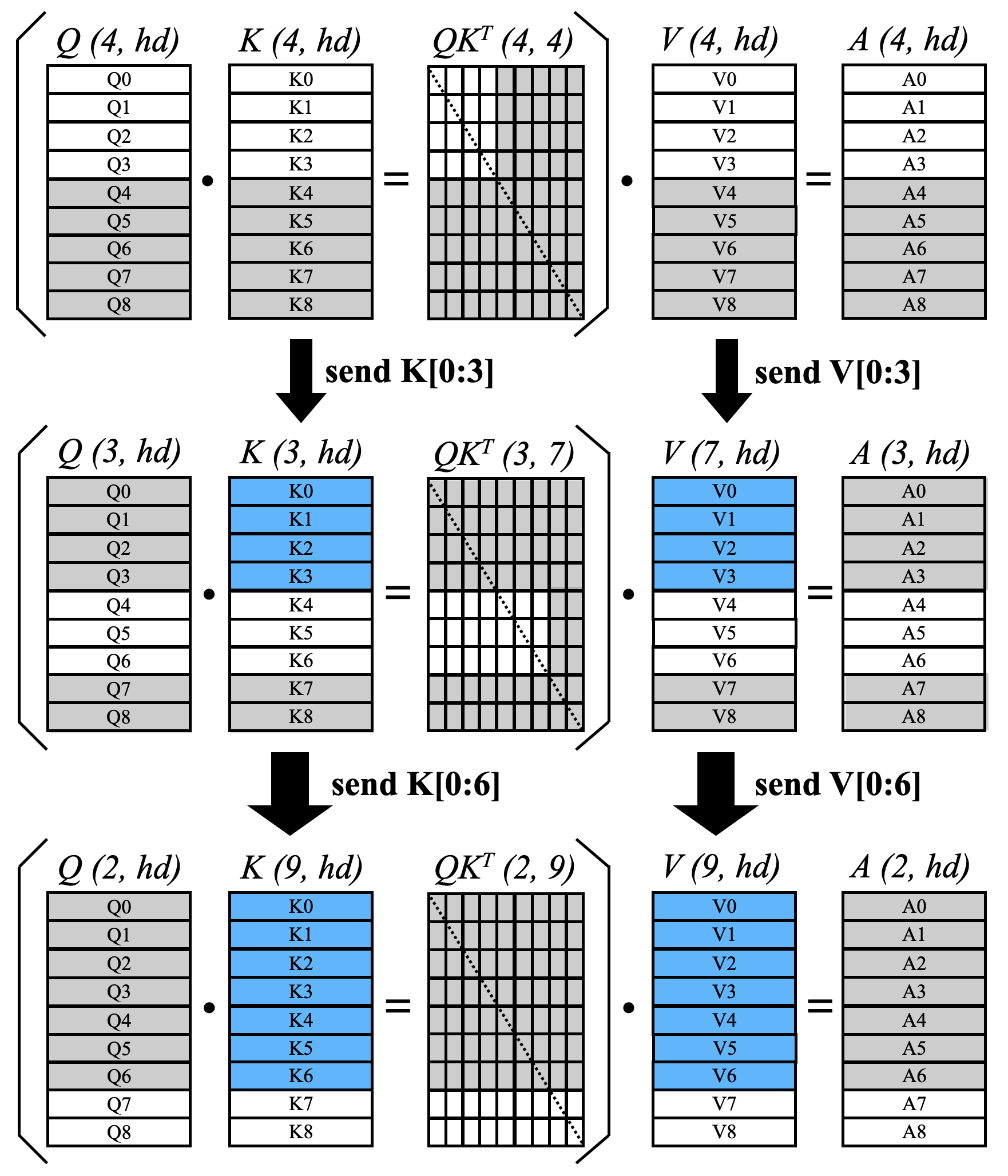}} 
 }

\vspace{-0.15 in}
\caption{ 
\prjname execution over  3 processes $p_{\{0,1,2\}}$ within a layer to compute attention map ($QK^T$) and final attention $A$: Each process will compute different amounts of $(Q,K,V)$ in (a), and the maximum amount for $QK^T$ is  \textbf{21 dot-products} on $p_1$   (in contrast to \textbf{27} from Fig.~\ref{dp::before_after} (b)). 
The locally computed $(K, V)$ are
passed down to the following processes as KV-cache using point-2-point one-way \textit{send}  (i.e., $p_0 \rightarrow p_1 \rightarrow  p_2$). Our communication is much cheaper than global \textit{all-gather} in Fig.~\ref{dp::before_after} (b), as the \textbf{traffic} incurred in \prjname is \textbf{22} (i.e., the number of blue rows in $(K,V)$), which is much lower than \textbf{36} from Fig.~\ref{dp::before_after}. 
}
\label{kvr::before_after}
\end{figure*}

\section{\prjname Overview}
In Fig.~\ref{kvr:overview}, the proposed \prjname is illustrated and compared against Tensor/sequence parallel inference (or \textbf{TSP}), which characterizes both tensor parallelism~\cite{shoeybi2020megatronlm,narayanan2021efficient} and sequence parallelization~\cite{li-etal-2023-sequence}.  
As   in Fig.~\ref{kvr:overview} (b), \prjname starts with uneven context partitioning for load-balancing. 
The existing TSP parallelizes the forward computation itself, but \prjname achieves parallel inference by utilizing multiple processes to populate KV-caches for the last process.
Therefore, unlike TSP where computation is symmetric and evenly distributed (thus no need to balance out context partitioning), 
\prjname needs good context partitioning to balance out the KV-cache amount from each compute process and to minimize TTFT.

Once partitioning is complete, each process will run each layer, conditioning on the KV-cache from its precedent process. In detail, the current process must wait for the required KV-cache to arrive from the precedent (i.e., notice the layer misalignment in Fig.~\ref{kvr:overview} (b)), forming a dependency chain via local peer-to-peer communication rather than global synchronization via \textit{all-gather}~\cite{thakur2005optimization}. 

We will first elaborate on how \prjname works inside each layer in terms of compute/communication saving in Section~\ref{kvr:layer}, and then discuss the context partitioning for load-balancing in \prjname in Section~\ref{kvr:hgs}. Finally, Section~\ref{kvr:impl} briefly discuss implementing \prjname.

\subsection{Forward Execution}
\label{kvr:layer}
The causal attention computation on a single process is shown in Fig.~\ref{llm_infer} (b), which is to be parallelized in this section.  For a given context, once $(Q, K, V)$ are computed,  $QK^T$ or attention map is computed for $A$. Although only the lower triangular part of $QK^T$ is needed due to the causality, the entire  $QK^T$ is commonly computed via dense matrix-matrix multiplication first, then a mask is added in general~\cite{HuggingFace-Transformers}, because no good mapping to BLAS-L3 exists or writing a custom kernel is expensive~\cite{CUBLAS}.

One SOTA way to enable parallel inference for LLM (e.g.,   GPT-3, Llama, and BLOOM),  would be to utilize tensor and sequence parallelizations~\cite{li-etal-2023-sequence,patel2023splitwise,shoeybi2020megatronlm,korthikanti2022reducing}, Tensor/sequence parallelization or \textbf{TSP} in  Fig.~\ref{dp::before_after} where the focus is on parallelizing the single process behavior from Fig.~\ref{llm_infer} (b).  
In TSP, for a given evenly partitioned context, $(Q,K,V)$ are independently computed on each process as in Fig.~\ref{dp::before_after} (a). Then, the collective operation \textit{all-gather} is performed to exchange $K$ and $V$ to all processes so that   $QK^T$  can be evenly distributed as shown in  Fig.~\ref{dp::before_after}  (b).
Although TSP faithfully follows the single process case in Fig.~\ref{llm_infer} (b), it does not take advantage of the causality in LLM inference.

In our \prjname, we start with a given yet unevenly partitioned context, and $(Q,K,V)$ are independently computed on each process as in Fig.~\ref{kvr::before_after} (a). Thus, each process  computes a different number of entries in $(Q,K,V)$. Then, \prjname simply populates the KV-cache from each process and hands over to the next process in chain, mimicking the extension phase in Fig.~\ref{llm_infer} (a). As a result, only the last process will have the full $(K, V)$, but still each process can output the  $A$ in the same shape as $Q$, driving the next layer. Since KV-cache itself is built upon the causality, \prjname can automatically 
minimize the computation of the upper triangle and reduce the number of dot-products for $QK^T$. For example, 27 dot-products are needed on all the processes in TSP as   in Fig.~\ref{dp::before_after} (b), but \prjname requires 21 (max out of $\{p_0:16, p_1:21, p_2:18\}$) as in Fig.~\ref{kvr::before_after} (b). 
This also highlights the motivation behind uneven context partitioning to 
minimize the largest $QK^T$ computation.

\prjname also removes the global synchronization points and reduces the total traffic volume exchanged among processes. \textit{ All-gather} operations in Fig.~\ref{dp::before_after} (b)  force all the processes to stop and secure the full $(K, V)$~\cite{thakur2005optimization}, while \prjname  shares only the local KV-cache with the next process via point-to-point \textit{send} operations. As a result, TSP in Fig.~\ref{dp::before_after} (a) requires to share 36 $(K,V)$ entries to get to the state in  Fig.~\ref{dp::before_after} (b), but \prjname only needs 22 to transit  to Fig.~\ref{kvr::before_after} (b). 
Such a dependency chain from KV-cache   introduces a longer wait time for the later processes, but   \prjname can outperform TSP even with such overheads.

In theory, with a sufficient number of parallel processes and a sufficiently   long user context (i.e., $QK^T$ dominates the runtime), \prjname can offer up to 2$\times$ speed up over TSP, because both total $QK^T$ computation and network traffic among processes in \prjname are half of those in TSP. It could be possible to handcraft a custom/expensive BLAS kernel for TSP to avoid over-computation. Even with a tailored custom kernel, however, the communication involved in TSP remains suboptimal as it still uses All-gather to exchange (K, V)  The proposed \prjname avoids both over-computation and wasted network traffic   seamlessly, by dual-purposing the LLM-specific KV-cache scheme (which already exists  for the extension phase).



Additionally, the same computational savings achieved with the custom GPU kernel, can also be applied to KVR. From Fig 5 (b), we can still see some wasted computation. Hence, a custom kernel would save such waste to further enhance the performance of KVR. Yet, the benefit from a custom kernel would diminish with more GPUs in parallel, as the nature of KV-cache allows our technique to approximate the unmasked lower triangle  more accurately with more processes, as illustrated in Figure 2 (b) and (d).


For simplicity, assume that a user context $C$ is even partitioned for \prjname and TSP over $p$ processes. Then, the total TSP traffic $\textbf{Net}_{tsp}$ can be written as follows:
\begin{align}
\textbf{Net}_{tsp}(C, p) &= p(p-1)\frac{C}{p}\\
&= (p-1)C
\label{kvr:net_tsp}
\end{align}
which is essentially the total number of $(K,V)$ entries from other processes. The total \prjname traffic $\textbf{Net}_{kvr}$ is the sum of the total KV-cache put into the network.
\begin{align}
\textbf{Net}_{kvr}(C, p) &= \frac{C}{p}+\frac{2C}{p}+\frac{3C}{p}+...\\
&= \sum_{i=1}^{p-1}\frac{iC}{p} = \frac{(p-1)}{2}C
\label{kvr:net_kvr}
\vspace{-0.1 in}
\end{align}

The 2$\times$ reduction is over the total computation and network traffic, not for each individual process. Therefore, it is critical to perform load-balancing to maximize the gain over TSP and minimize TTFT, and \prjname accomplishes it by context-level load-balancing in Section~\ref{kvr:hgs}.

\subsection{Context-Level Load-Balancing}
\label{kvr:hgs}
As discussed in Section~\ref{motiv}, \prjname needs load-balancing for low TTFT.
We propose running off-line search for the best partitioning, and then store the results in a partitioning lookup table. 
For example, we pre-compute the optimal partitioning of user contexts at various lengths  for a given number of processes off-line by measuring TTFT on the target hardware, and then contribute the search results to a lookup table. During inference, we can predict the best partitioning by interpolating the two nearest known entries in the lookup table. For the example of 10k prompt, we can interpolate from the known partitioning configurations from 8k and 12k in the lookup table.

Finding the best partitioning configuration for a given user context, although one-time off-line overhead, can be expensive. Hence, we propose  a hierarchical grid search for acceleration. 
From the nature of \prjname, it is straightforward to see that finding the TTFT-optimal partitioning  has two conflicting objectives.

\begin{itemize}  
\vspace{-0.1 in}    
   \item The partitions for the earlier processes must be small, otherwise the later processes will wait too long for the earlier ones  to populate KV-caches and send them over.
   \item The partitions for the later processes need to be small, otherwise the later processes will the bottleneck in generating the first token.
   \vspace{-0.1 in}    
\end{itemize}

\begin{figure}[!t]
\centering
   \centering
\mbox{
\hspace{-0.1in}
		\subfigure[2 partitions ]
		{\includegraphics[width=0.8 in]{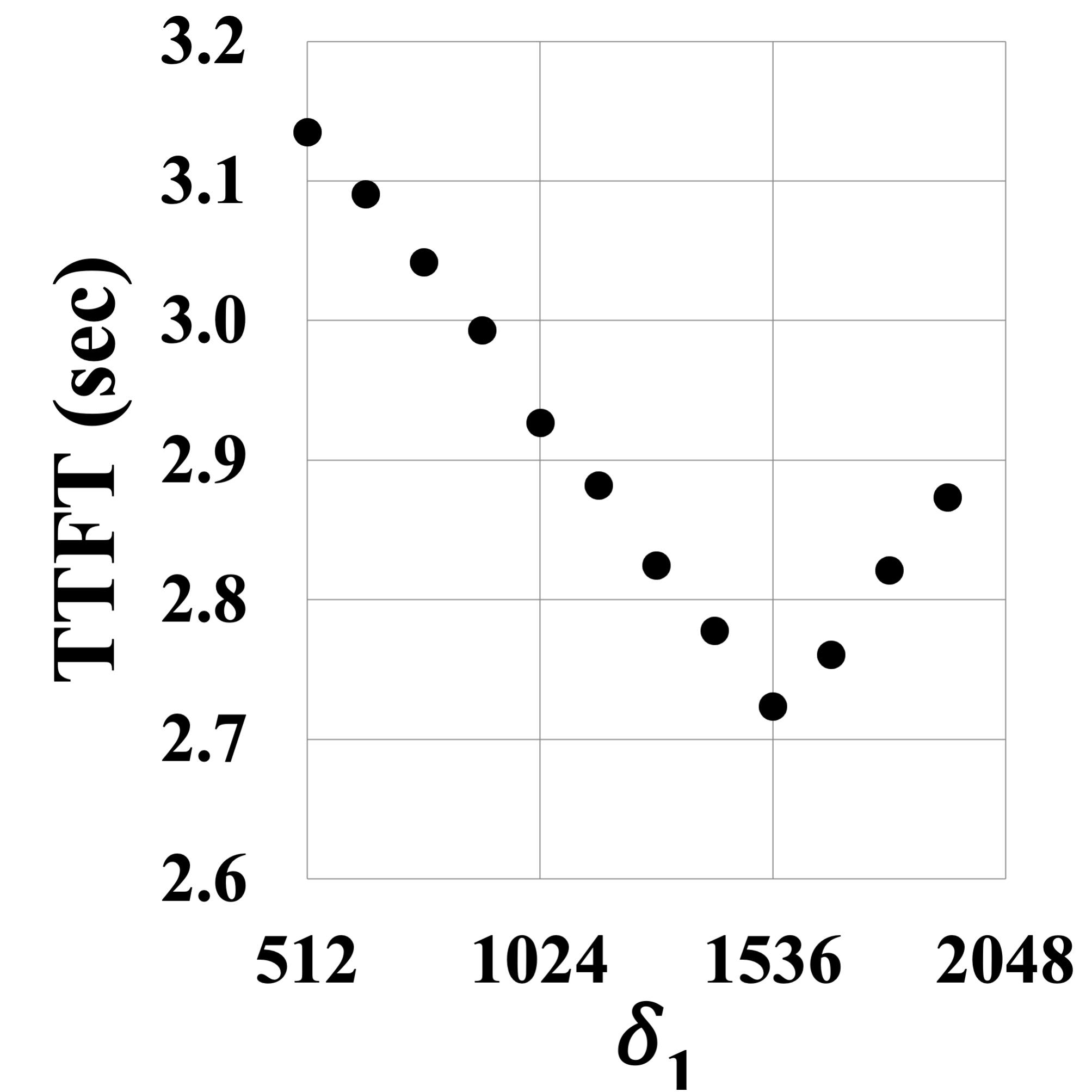}} 
		\subfigure[stride 8]
		{\includegraphics[width=0.8 in]{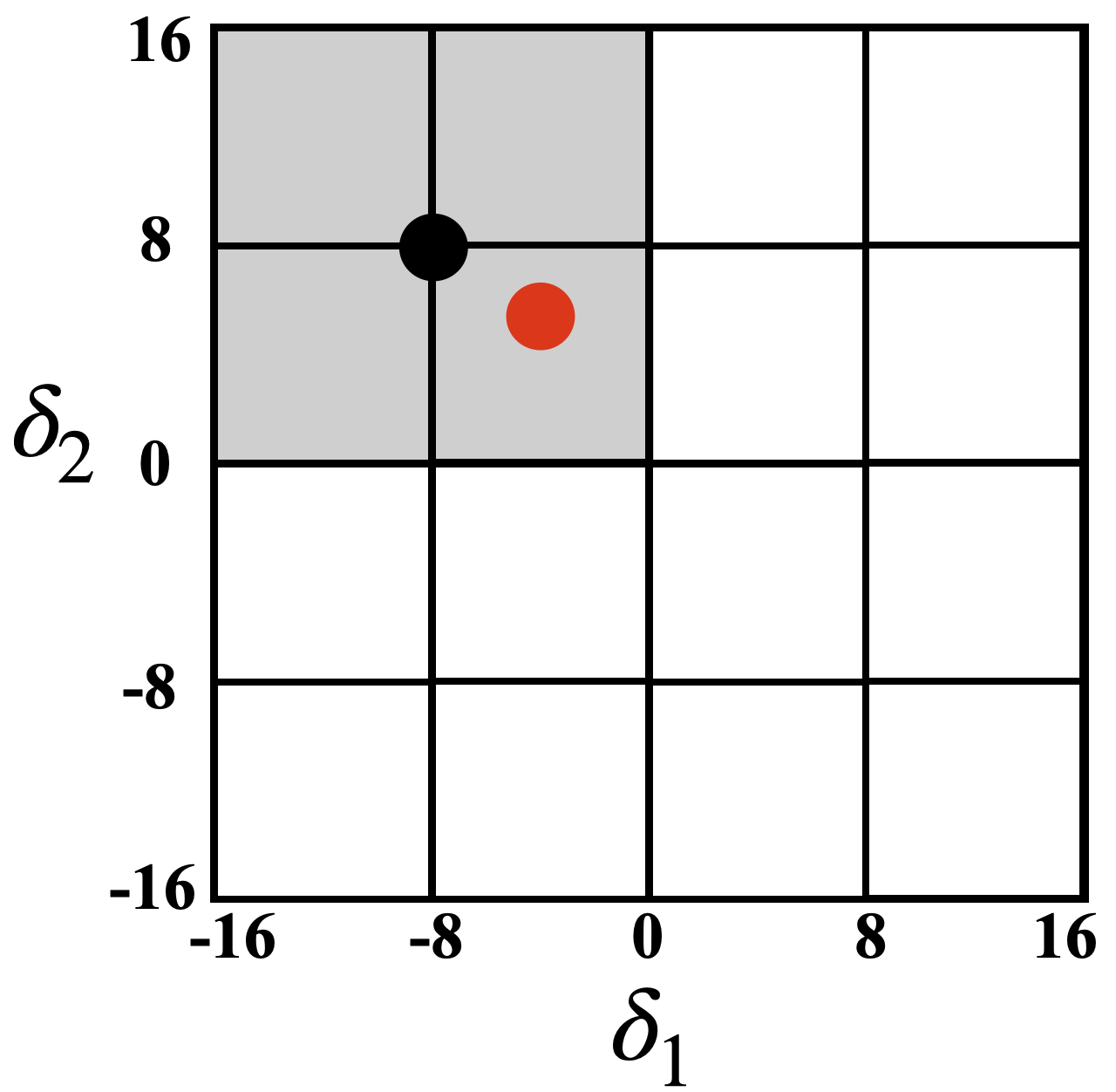}}  

		\subfigure[stride 4]
		{\includegraphics[width=0.8 in]{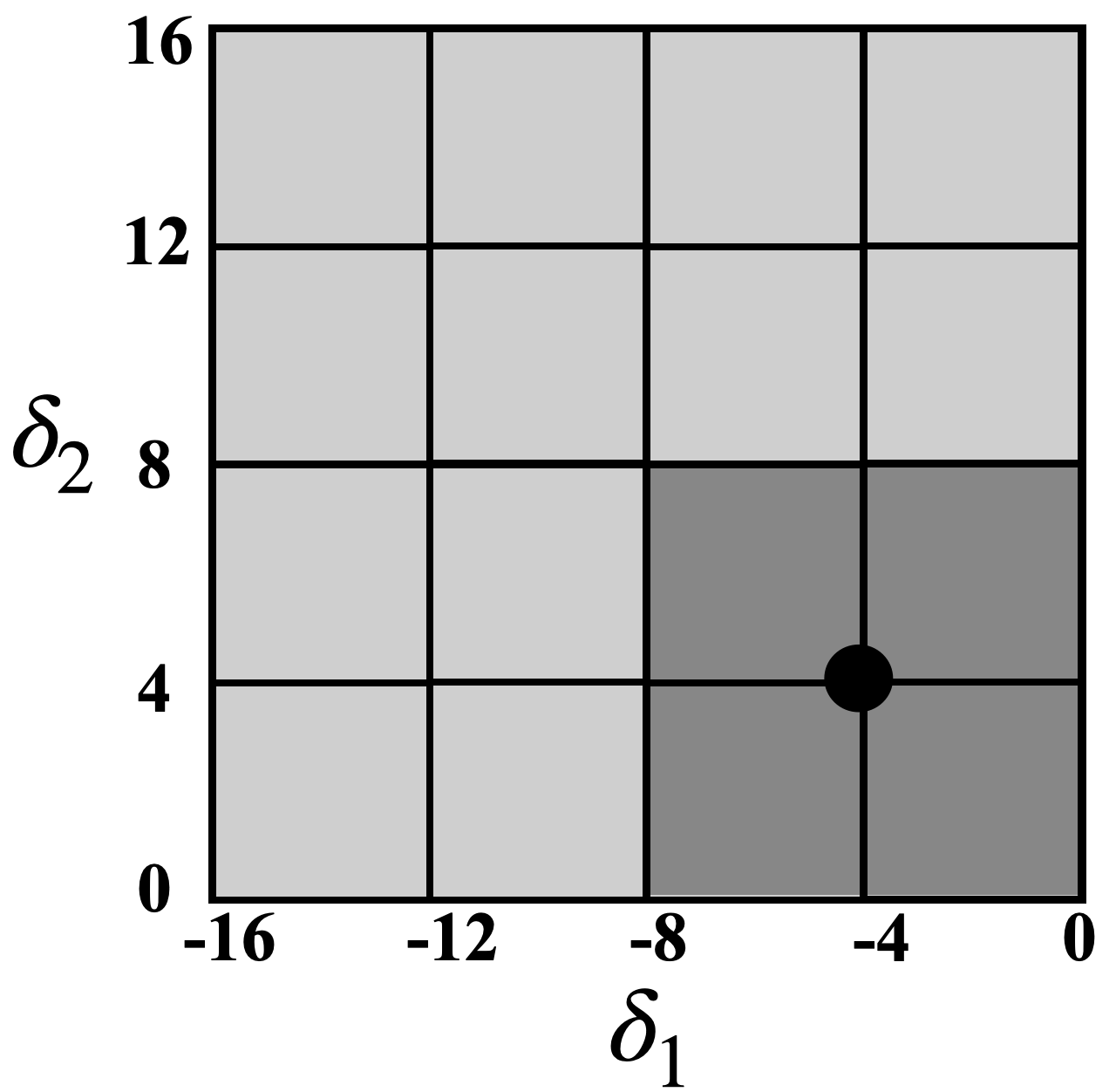}} 

		\subfigure[stride 2]
		{\includegraphics[width=0.8 in]{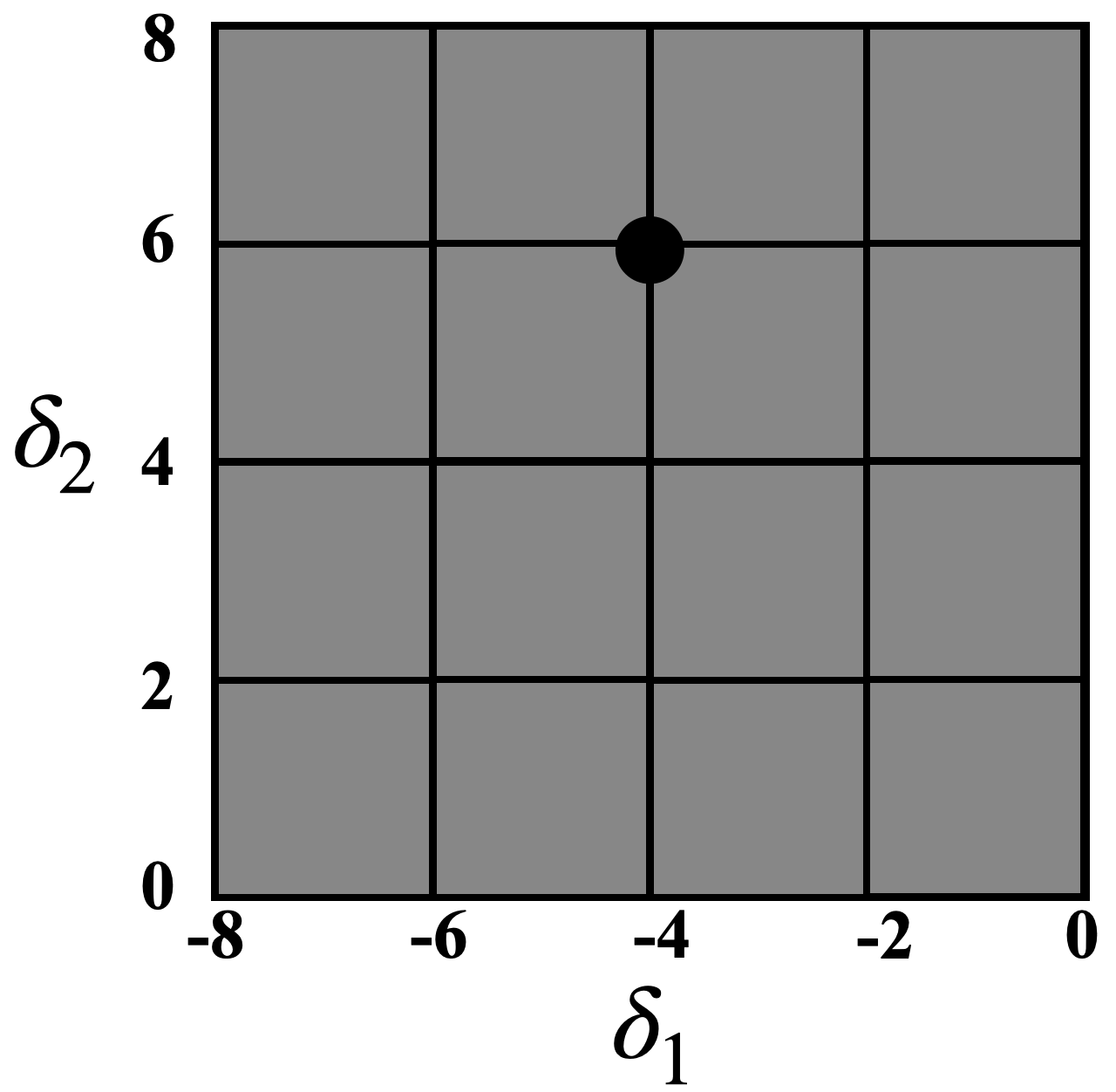}}   
 }  
  \vspace{-0.15 in}
\caption{ 
Finding two partitions can be done quickly by binary search as shown in (a) where $\delta_1$ is the variable for additional context the $p_0$.
We can extend such binary search into hierarchical grid search for multiple processes as in (b, c, d)  for the example of 
$C[0, 32+\delta_1, 64+\delta_2, 96]$.}
\label{kvr:partition}
     \vspace{-0.1 in}
\end{figure}

For two processes, we can use a binary search to find out the best partitioning. Fig.~\ref{kvr:partition} (a) shows how TTFT changes as we grow the partition for the $p_0$ for a 16k context where the partitioning is $C[0, 8192+\delta_1, 16384]$. As $\delta_1$ grows, it bottoms at the partition of $[0, 9728, 16384]$ (i.e., $\delta_1=1536$, thus $p_0$ takes C[0:9728] and $p_1$ takes in C[9728:16384]).

By generalizing the binary search into a hierarchical grid search for multiple processes~\cite{zhang2022novel}, we can find a high-quality partitioning fast for a given user context length. Figs.~\ref{kvr:partition} (b-d) depict the proposed search process for a user context length of 96 over 4 processes, which is to find the optimal  $(\delta_1, \delta_2)$ for the partitioning of $C[0, 32+\delta_1, 64+\delta_2, 96]$. In the first level, we set the search stride as 8 and  measure the TTFTs on each grid. Once we find the best performing $(\delta_1, \delta_2)$ pair, we limit the search to the gray grid and reduce the search stride to 4 to perform another scan as in Fig.~\ref{kvr:partition} (c). We repeat the same process recursively until the minimum stride is applied, leading to the final search as in Fig.~\ref{kvr:partition} (d). The best partitioning is then $[0, 28, 70, 96]$ and marked as a red dot in Fig.~\ref{kvr:partition} (b).

A comprehensive partitioning lookup table will enable efficient  partitioning as in Fig.~\ref{kvr:overview} (b) for effective load-balancing. For a given user context, we will interpolate and predict the best partitioning from two closest entries. Therefore, having a dense and large table would be advantageous at the cost of one-time  search overhead. Our results also show that even at the 4k intervals between entries, the predicted partitioning can yield excellent TTFT (see Fig.~\ref{ttft:llama7b_part}).

\begin{figure}[!t]
\centering
   \centering
\mbox{

		\subfigure[Without \prjname.]
		{\includegraphics[width=1.34 in]{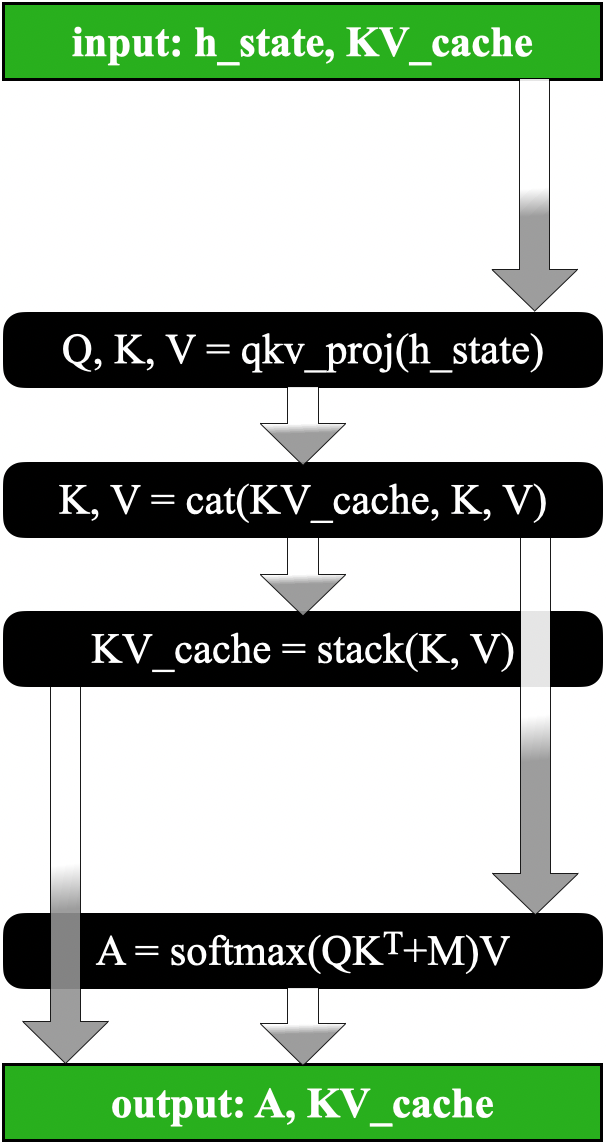}} 
 
  \hspace{0.6 cm}
		\subfigure[With \prjname. ]
		{\includegraphics[width=1.34 in]{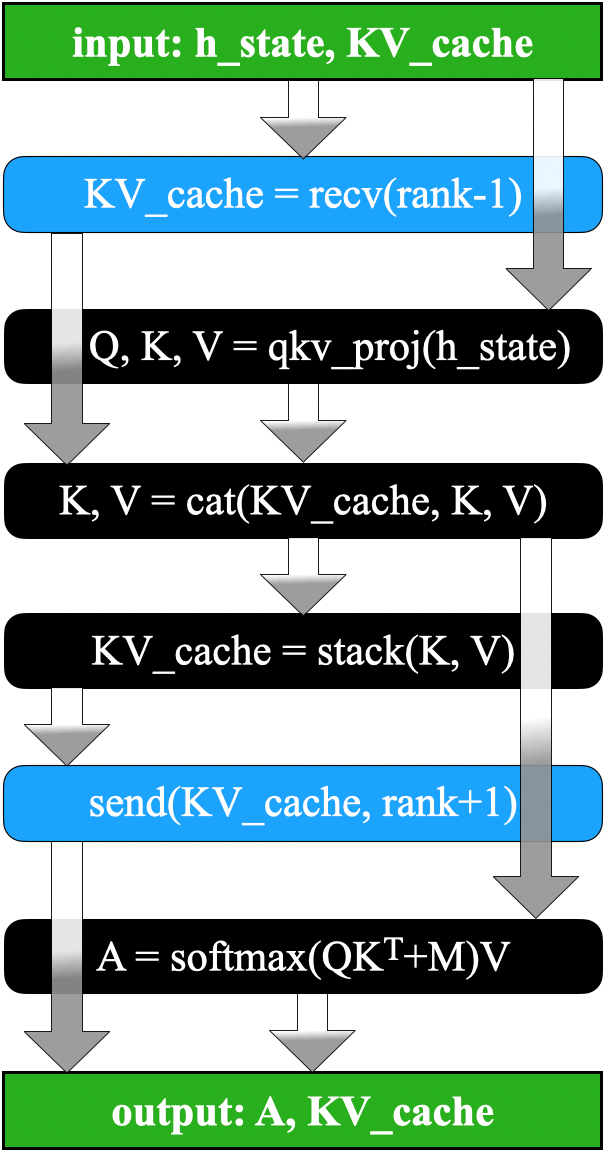}} 
 }  
\vspace{-0.1 in}
 \caption{ 
\prjname can be easily implemented in LLM with existing KV-cache support (e.g., most of public LLMs) by simply inserting \textit{recv/send} operations (in the blue boxes). Note that $M$ is the causality mask.}
\vspace{-0.12 in}
\label{kvr:implm}
  
\end{figure}

\begin{figure*}[!t]
\centering
   \centering
\mbox{
\hspace{-0.1 in}
		\subfigure[\textbf{TSP} hits out-of-memory error for 16k.]
		{\includegraphics[width=2.25 in]{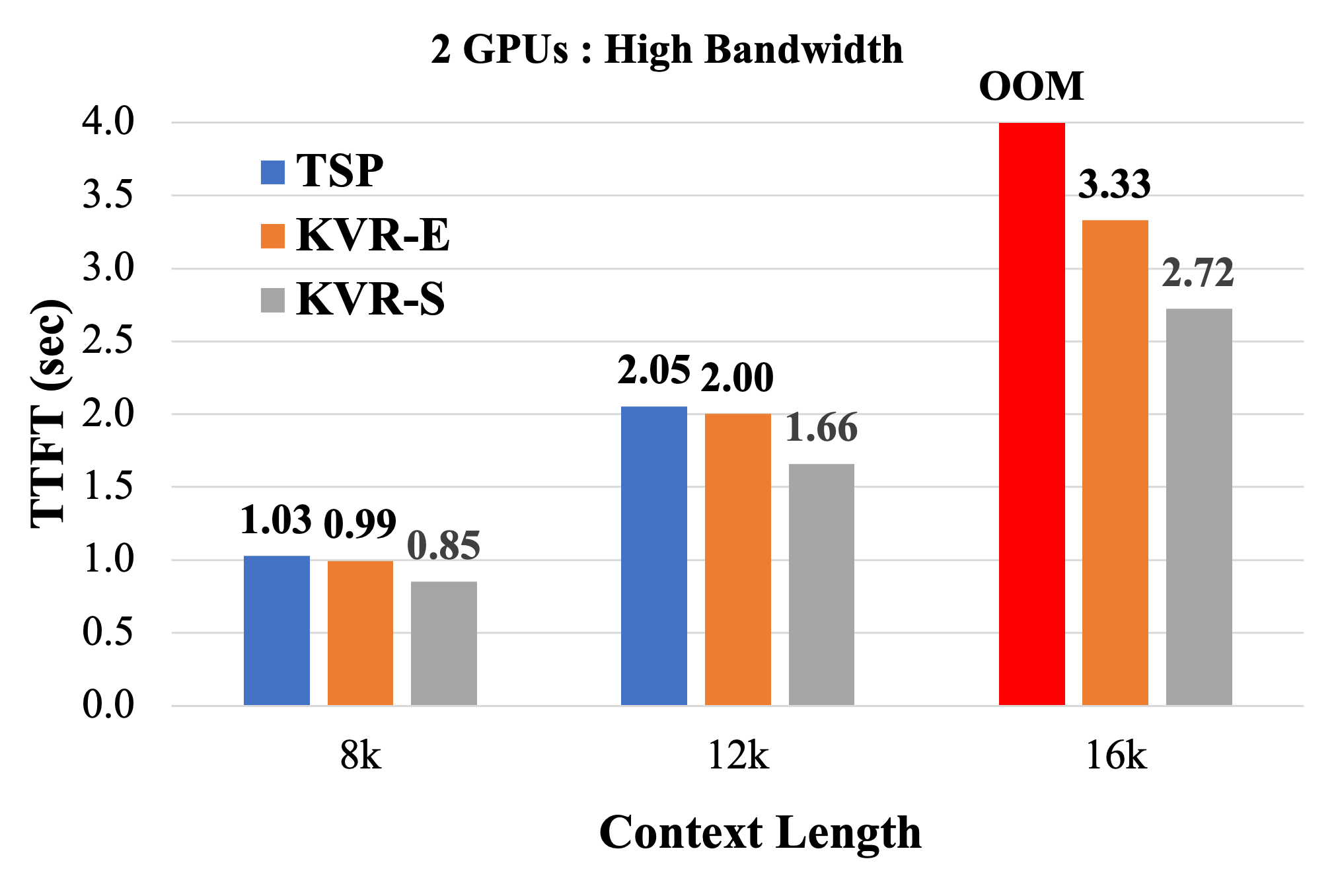}} 
 
		\subfigure[\textbf{KVR-S} is 1.42$\times$ faster for 12k   than \textbf{TSP}. ]
		{\includegraphics[width=2.25 in]{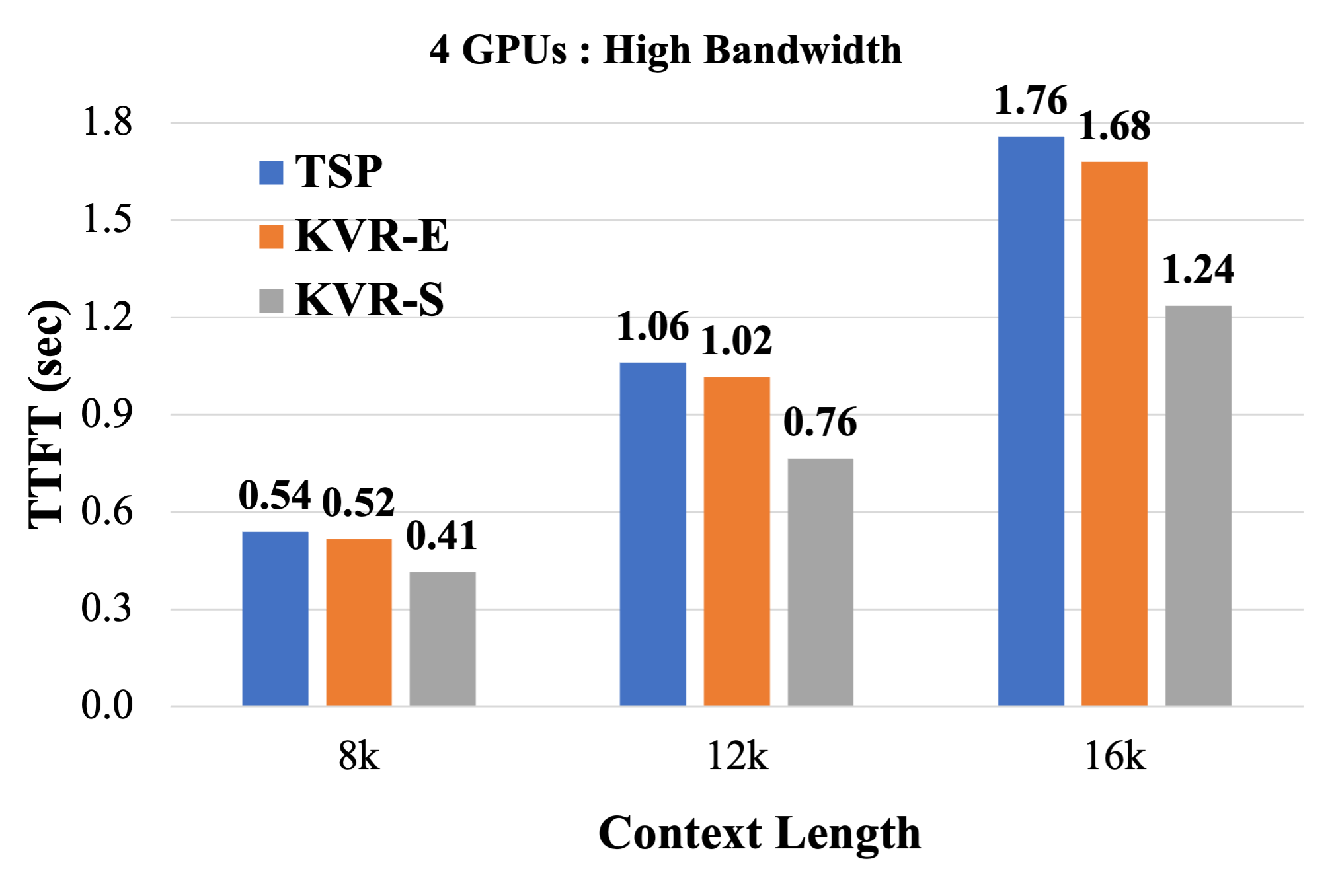}} 


		\subfigure[\textbf{KVR-S} is 1.41$\times$ faster for 16k than \textbf{TSP}.]
		{\includegraphics[width=2.25 in]{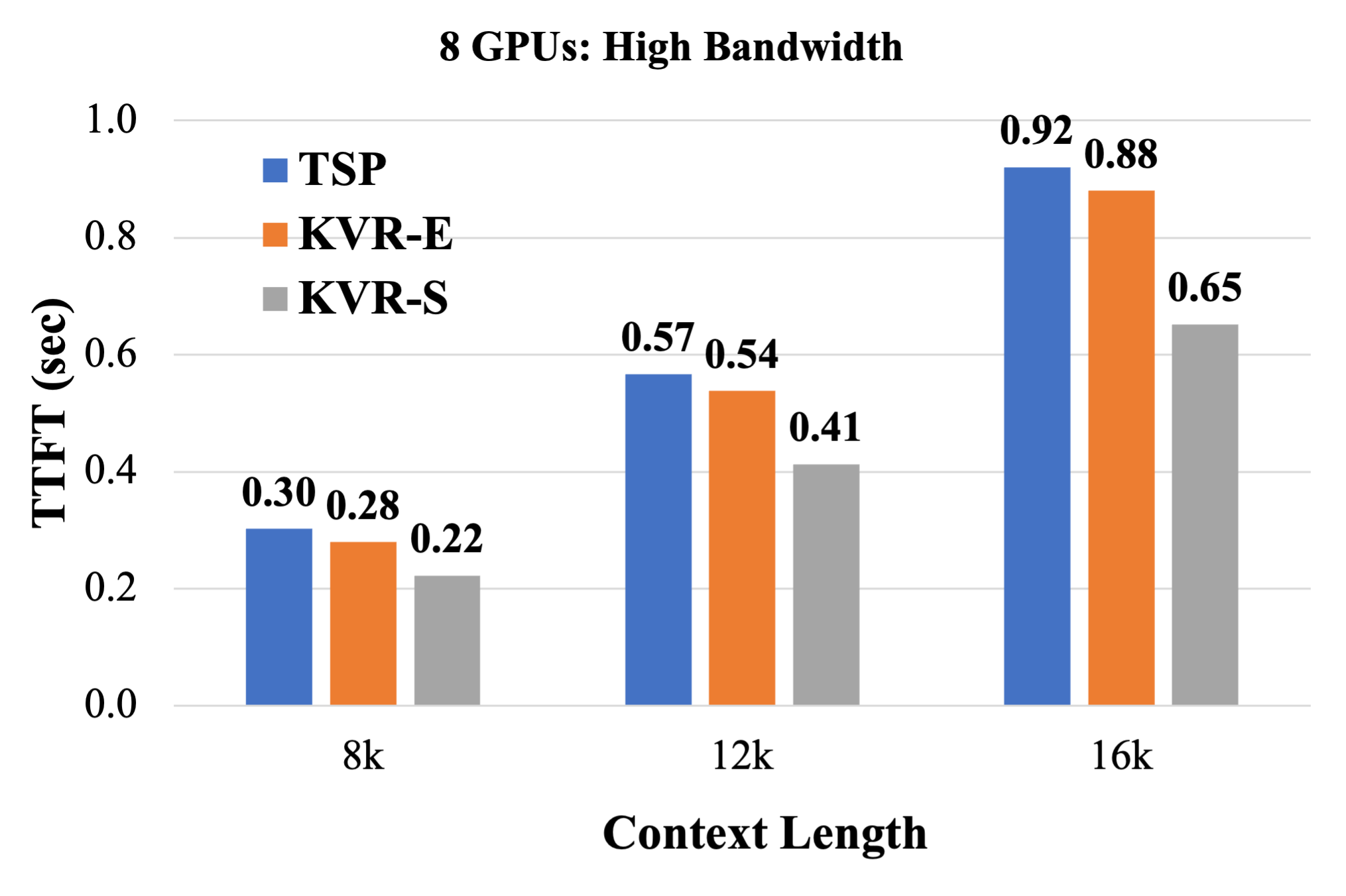}} 
 
 }  

\mbox{
\hspace{-0.1 in}
		\subfigure[\textbf{KVR-S} shows the best scalability.]
		{\includegraphics[width=2.25 in]{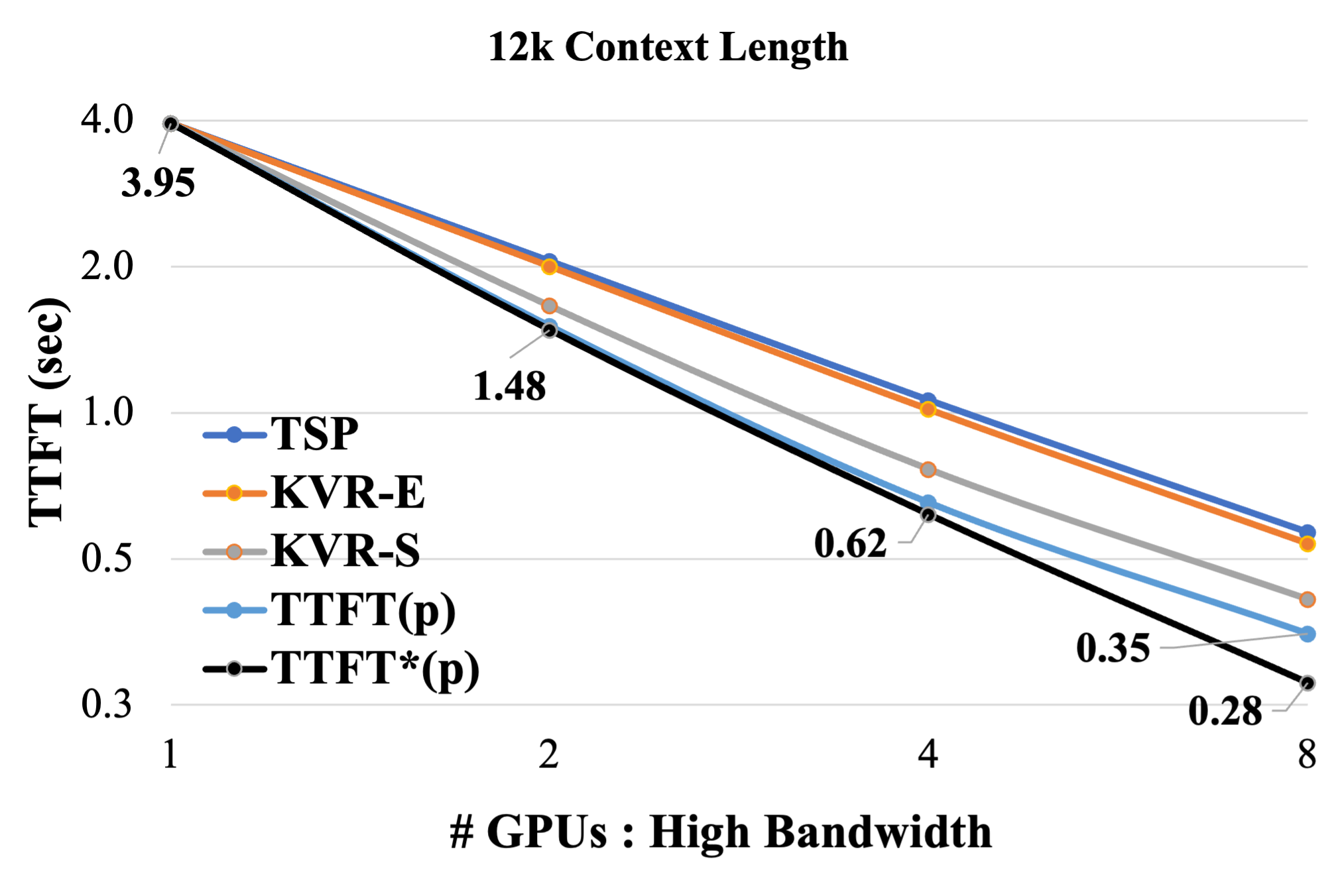}}
  
  \subfigure[\textbf{KVR-S} is 1.79$\times$ faster for 8k than \textbf{TSP}.]
		{\includegraphics[width=2.25 in]{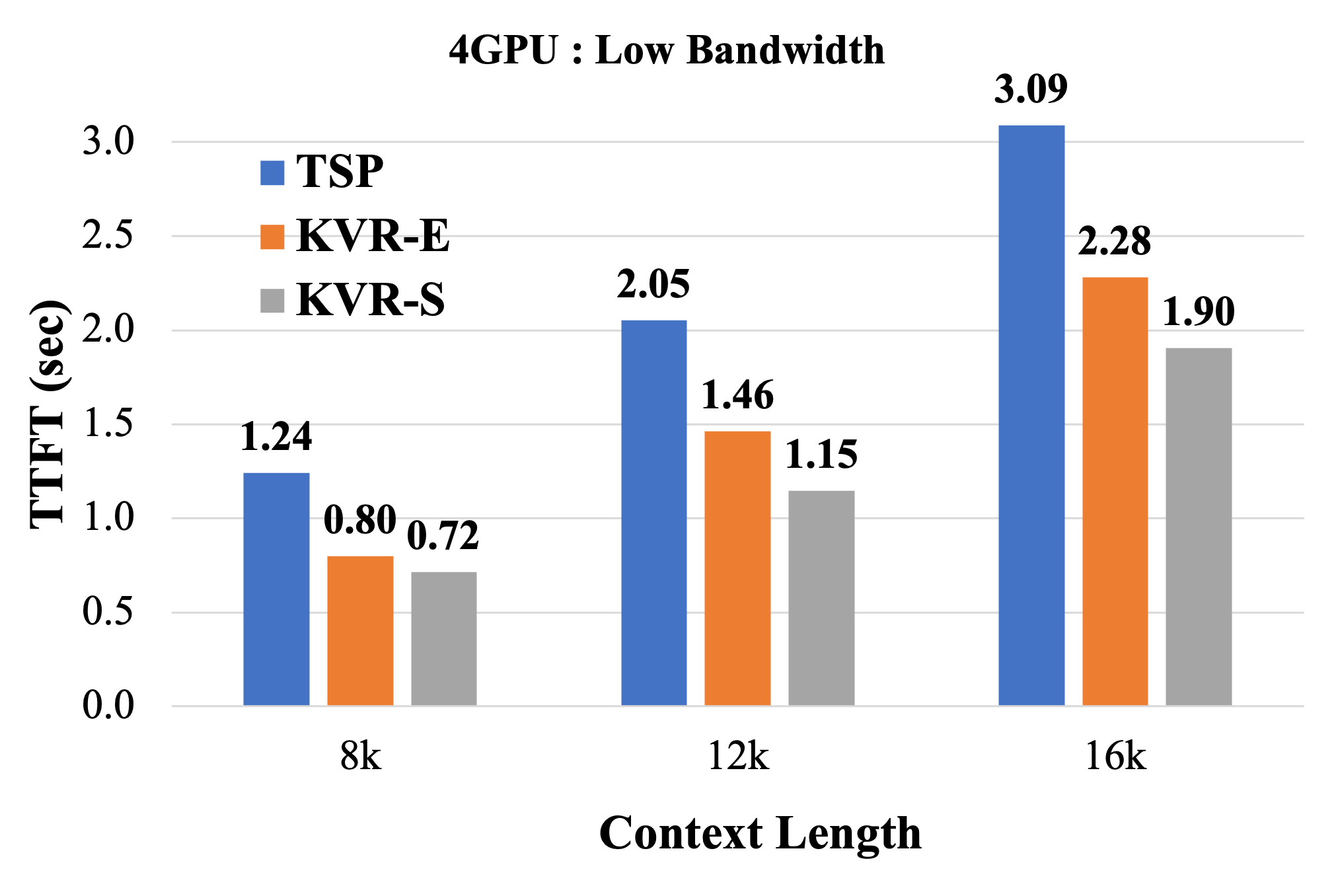}} 

		\subfigure[\textbf{KVR-S} is 1.57$\times$ faster for 16k than \textbf{TSP}. ]
		{\includegraphics[width=2.25 in]{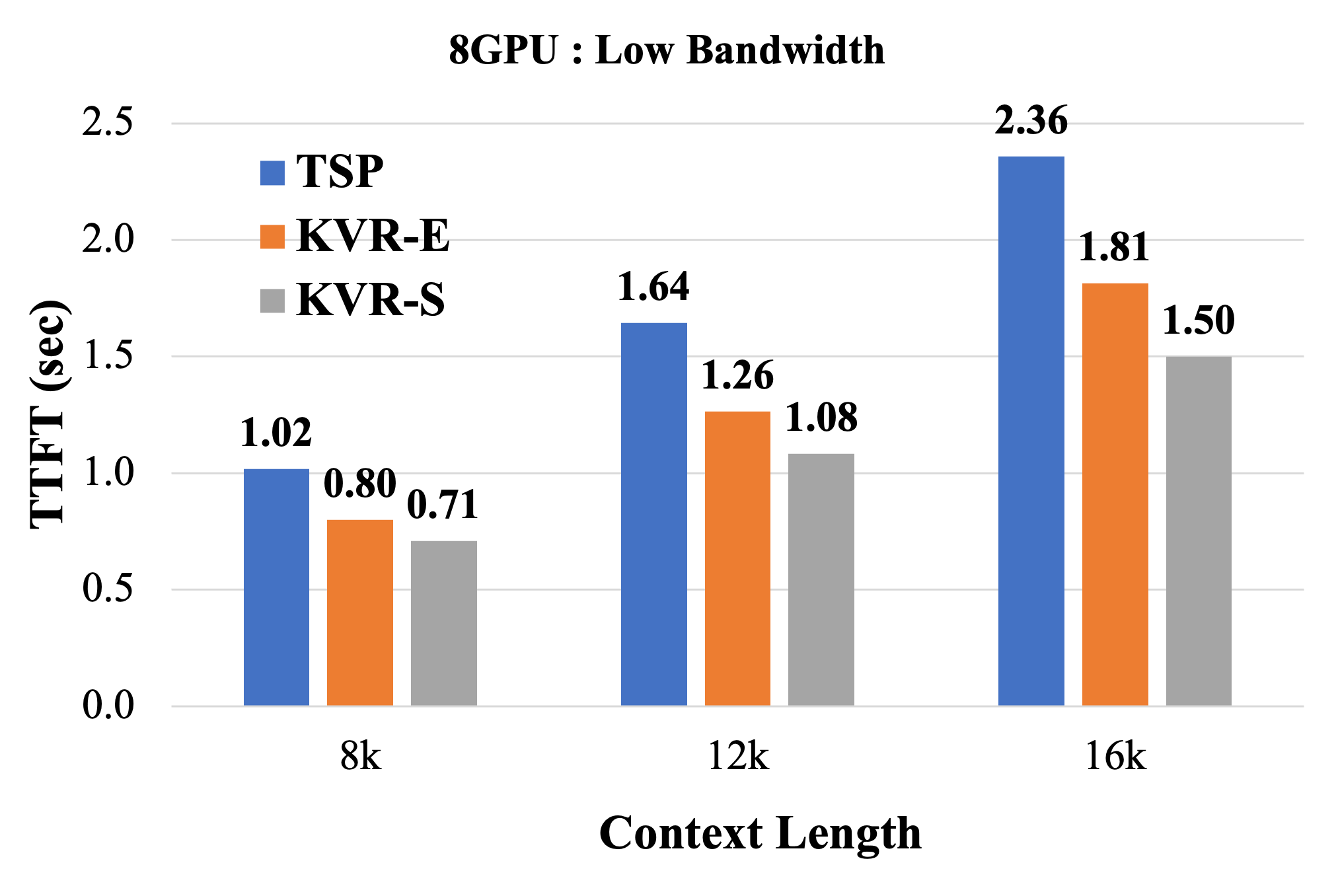}} 

 
 }  
 \vspace{-0.1 in}
\caption{\textbf{Llama 7B:} while \textbf{KVR-E} already outperforms \textbf{TSP} in all the test cases,  
\textbf{KVR-S} further accelerates by 1.42$\times$ over \textbf{TSP} as in (a-c) with 300GB/s network.
The speedups from \textbf{KVR-E} and \textbf{KVR-S}  are even higher with 10GB/s network as in (e, f): 1.55$\times$ (4 GPUs and 8k) and  1.79$\times$ (4 GPUs and 12k) over \textbf{TSP}, respectively.
\textbf{KVR-S} is the closest to the scalability lower bounds as in (d).}
\label{ttft:result_llama7b}
\end{figure*}

\begin{figure*}[!t]
\centering
   \centering

\mbox{
\hspace{-0.1 in}
		\subfigure[\textbf{KVR-S} is 1.26$\times$ faster for 4k than \textbf{TSP}.]
		{\includegraphics[width=2.25 in]{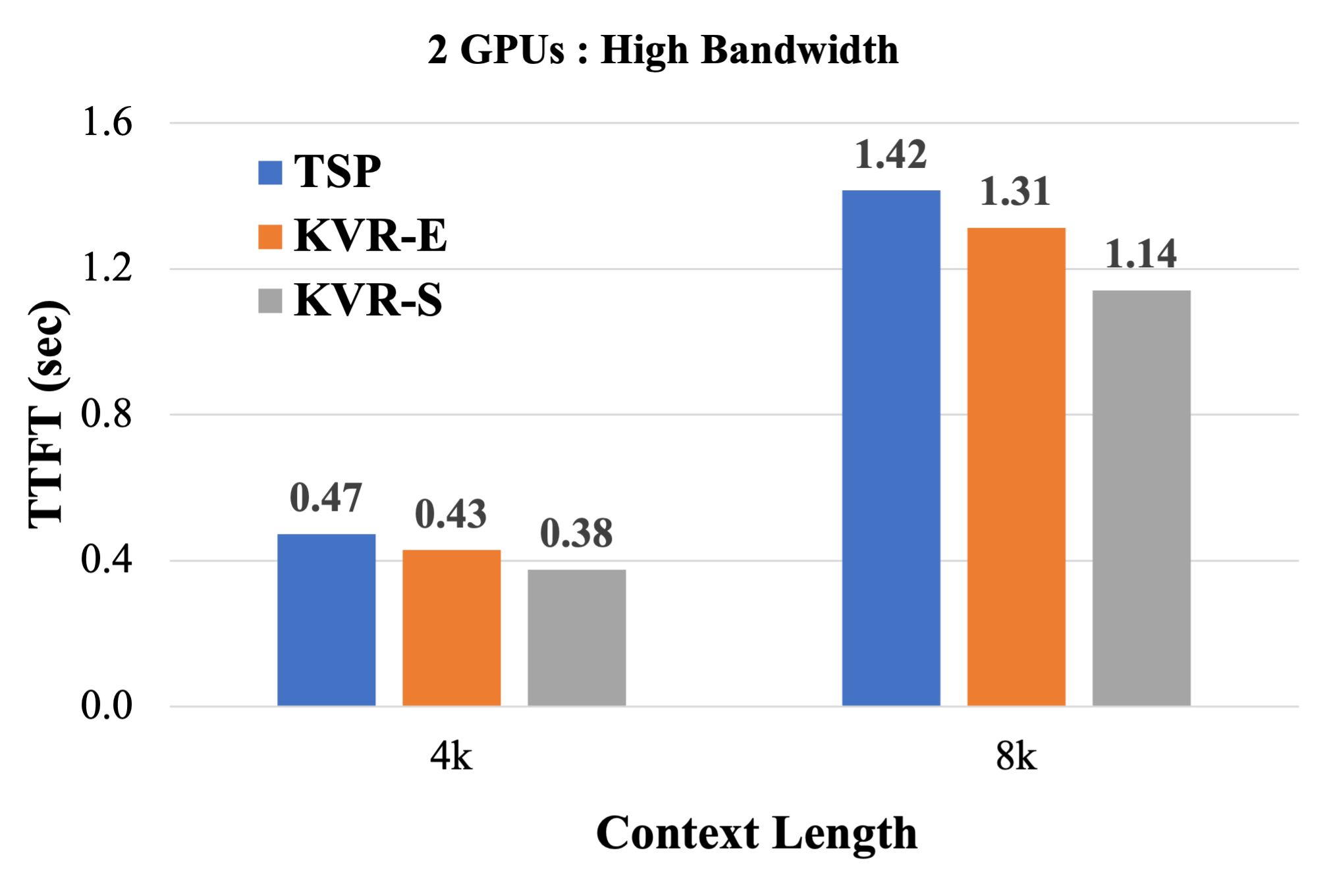}}
  
  \subfigure[\textbf{KVR-S} is 1.46$\times$ faster for 8k than \textbf{TSP}.]
		{\includegraphics[width=2.25 in]{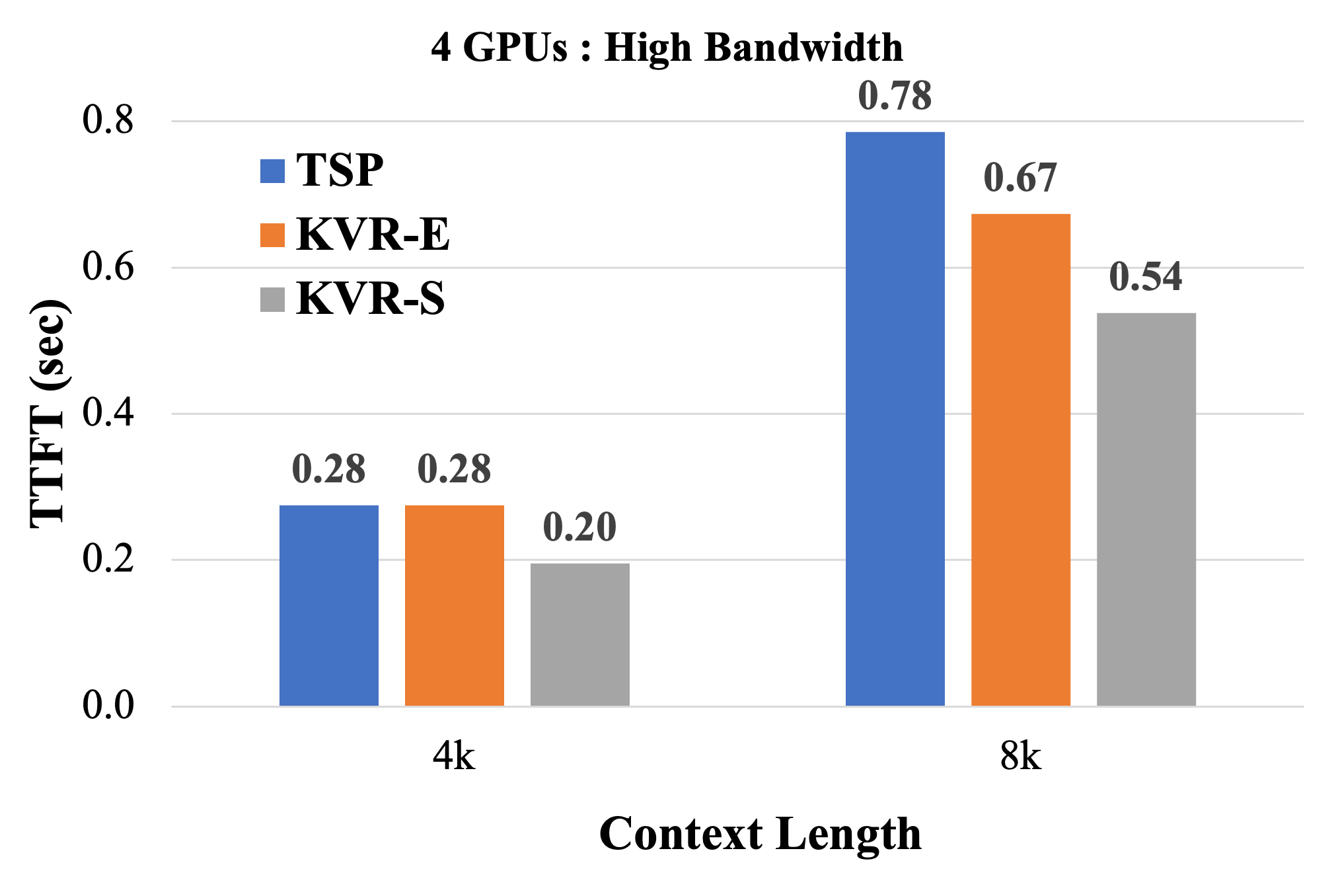}} 

		\subfigure[\textbf{KVR-S} is 1.63$\times$ faster for 8k than \textbf{TSP}. ]
		{\includegraphics[width=2.25 in]{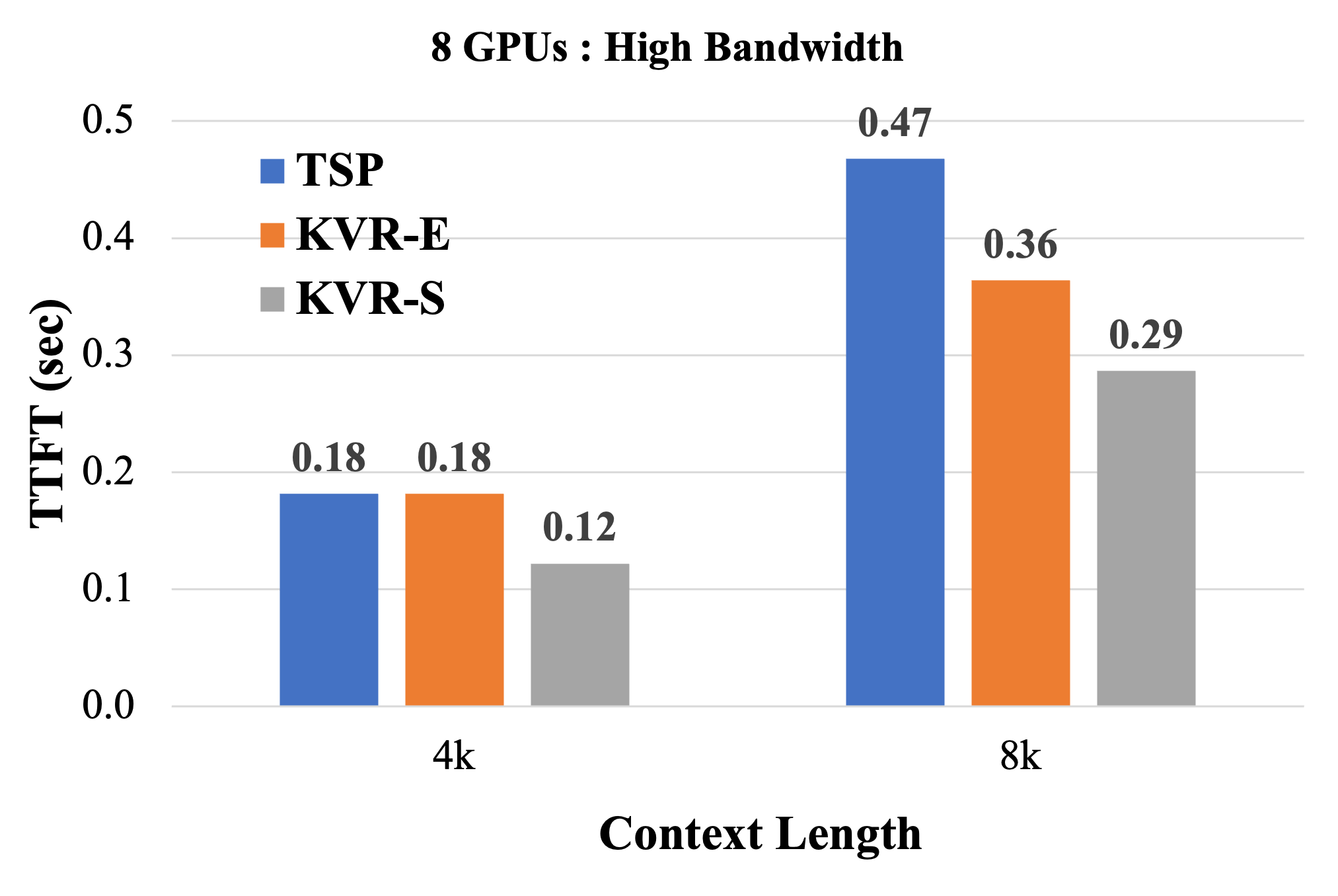}} 

 
 }  
	
 \vspace{-0.1 in}	

\caption{ \textbf{Falcon 7B:}  
\textbf{KVR-S} offers up to 1.63$\times$  speedup over \textbf{TSP}. Since 4k context length is relatively short, the benefit from \textbf{KVR-E} is canceled out by the overhead from KV-cache waiting time and unbalanced attention compute as in (b-c). However, with load-balancing,  \textbf{KVR-S} delivers 1.37$\times$ and 1.47$\times$ speedup over \textbf{TSP} with 4 and 8 GPUs, respectively, which emphasizes the context-level load-balancing.}
\label{ttft:result_falcon7b}
\end{figure*}

\begin{figure*}[!t]
\centering
   \centering
\mbox{

		\subfigure[The  partitioning breakdowns  found by hierarchical grid search for     \textbf{KVR-S}   in Figs.~\ref{ttft:result_llama7b} (a-c).]
		{\includegraphics[width=2.3 in]{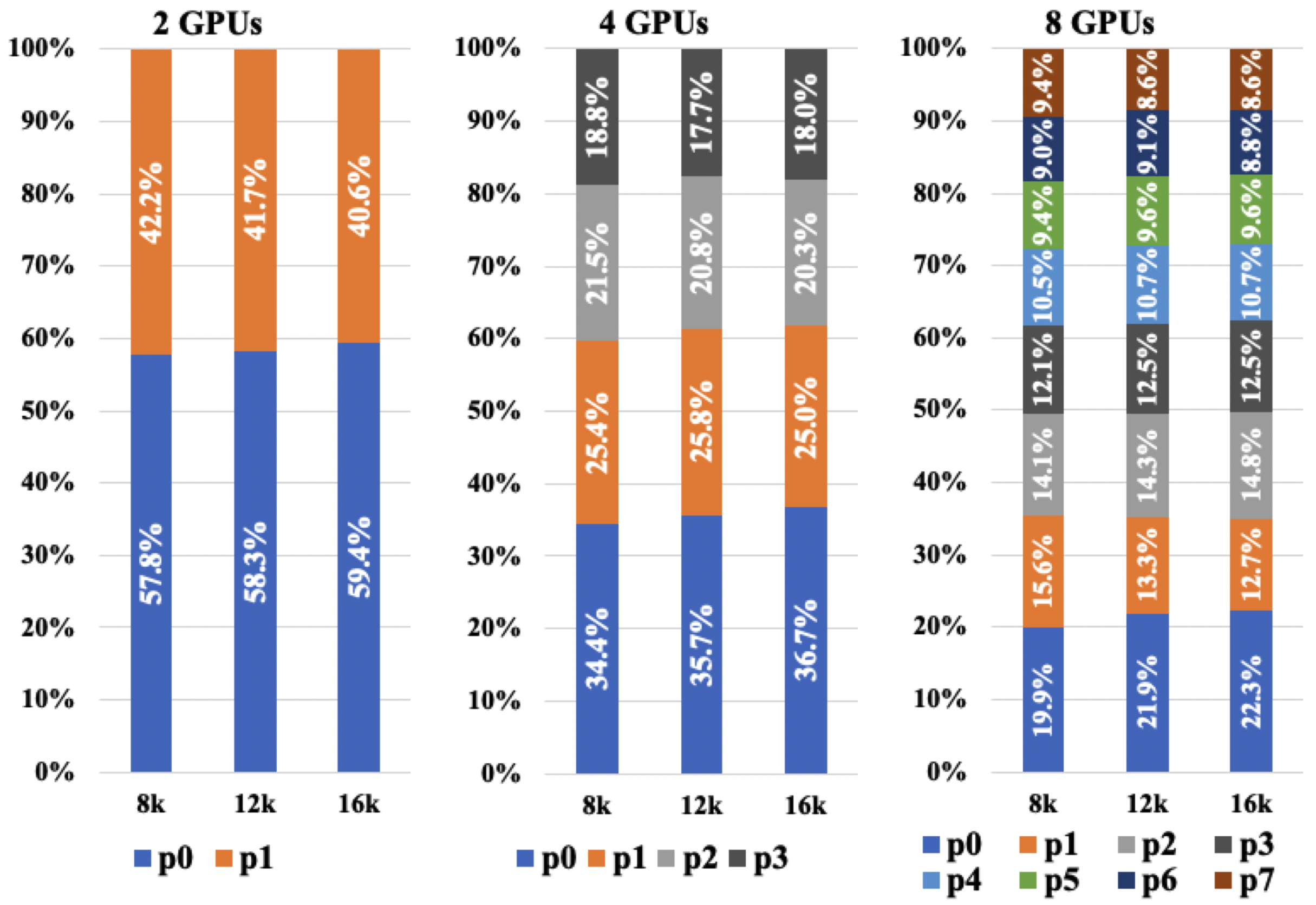}} 
  \hspace{0.05in}

		\subfigure[\textbf{KVR-P} with predicted partitions is only 1.1\% worse than \textbf{KVR-S}.]
		{\includegraphics[width=2.2 in]{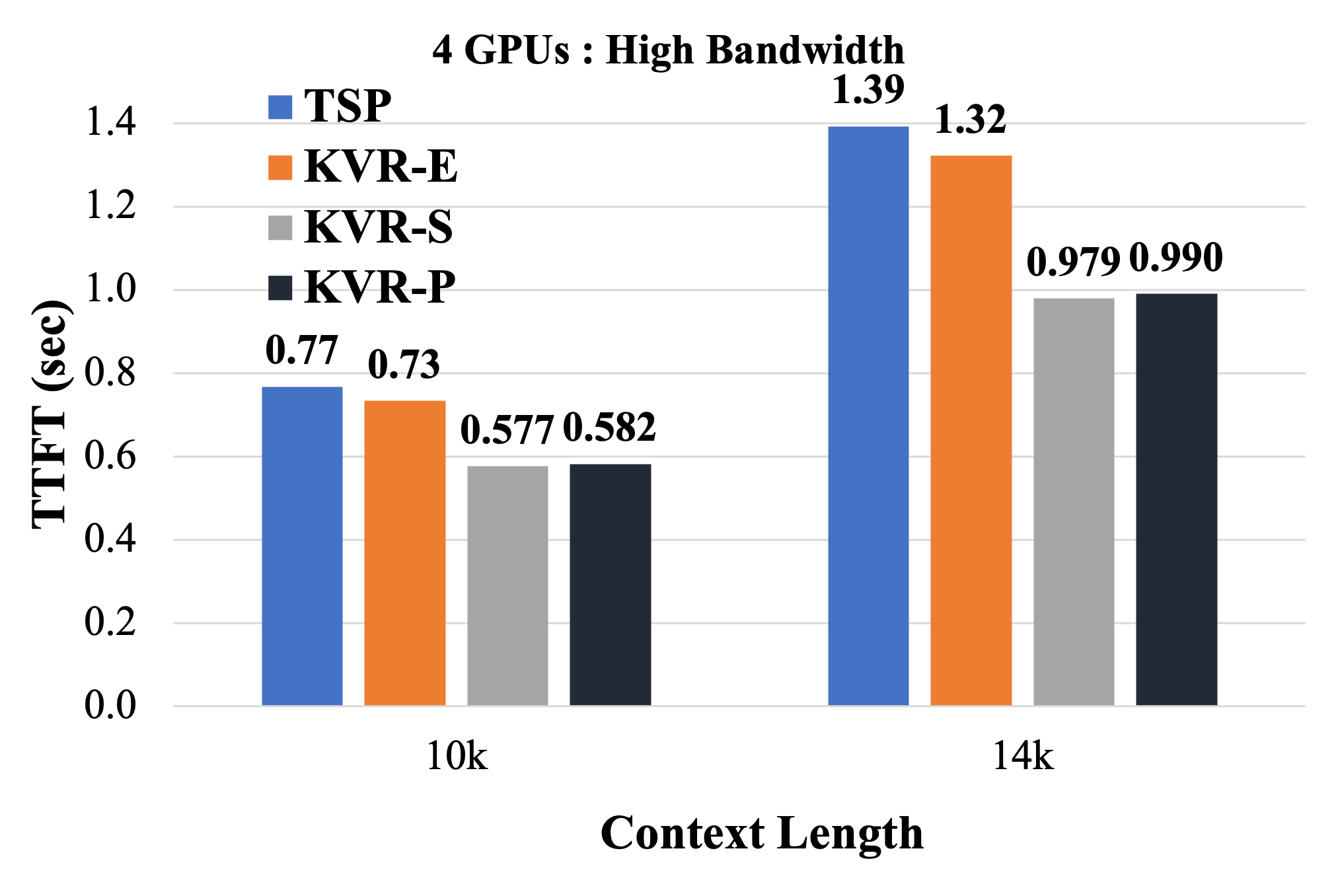}} 
\hspace{0.05in}
 
		\subfigure[\textbf{KVR-P} with predicted partitions is only 1.3\% worse than \textbf{KVR-S}.]
		{\includegraphics[width=2.2 in]{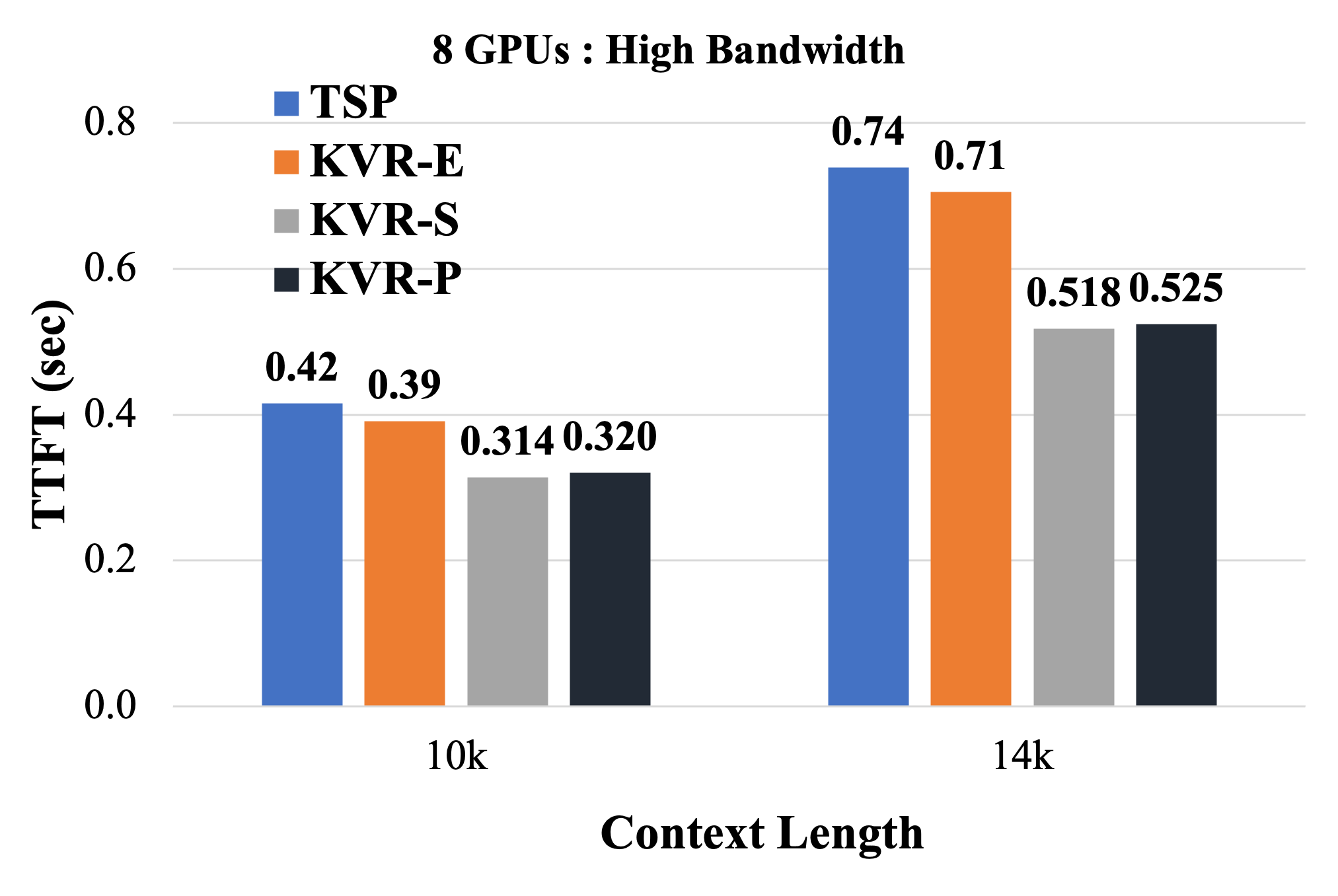}} 
 }

\vspace{-0.10 in}
\caption{ 
\textbf{KVR-P} with predicted partitions
for 10k and 12k contexts interpolated from the searched breakdowns for Figs.~\ref{ttft:result_llama7b} (a-c) has negligible TTFT degradations from \textbf{KVR-S}, supporting the fact that the proposed lookup method for context partitioning  is effective.}
\label{ttft:llama7b_part}
    \vspace{-0.2 in}/
\end{figure*}



\subsection{Implementation}
\label{kvr:impl}
Since \prjname dual-purposes the KV-cache interface, which exists in most LLM implementations for faster subsequent token generations during the extension phase   in Fig.~\ref{llm_infer} (a)~\cite{HuggingFace-Transformers}, \prjname is easy to implement. Fig.~\ref{kvr:implm} shows the  pseudocode/computational graph without and with \prjname. Note that KV-cache is already in the input argument to the attention block.
The only additions are two parts in the blue boxes: \textbf{a)} overwrite the KV-cache by receiving it from $p_{i-1}$ before concatenating it with the local $(K, V)$, and \textbf{b)} forward the updated KV-cache to $p_{i+1}$ right after concatenation.
We can make both \textit{recv} and \textit{send} asynchronous  calls by overlapping  with \textit{qkv\_proj} and \textit{softmax} respectively, thanks to the nature of point-to-point connections.
More details on the implementation and examples can be found in Appendix~\ref{kvr:pseudo_code}.

Both TSP and \prjname require to have tensors in the contiguous memory space for efficient network communication, which is then about KV-cache  for \prjname: if KV-cache is physically fragmented, costly extra memory copy will be necessary. Therefore, the KV-cache management such as vLLM
~\cite{kwon2023efficient,vllm} needs to support contiguous physical memory allocation during the prompt phase to work seamlessly  with \prjname.



\begin{figure*}[!t]
\centering
   \centering
\mbox{
		\subfigure[On a noisy network, \textbf{KVR-S} outperforms \textbf{TSP} even more (42.2\% vs. 43.4\%) than the quiet case in Fig.~\ref{ttft:result_llama7b} (b).]
		{\includegraphics[width=2.2 in]{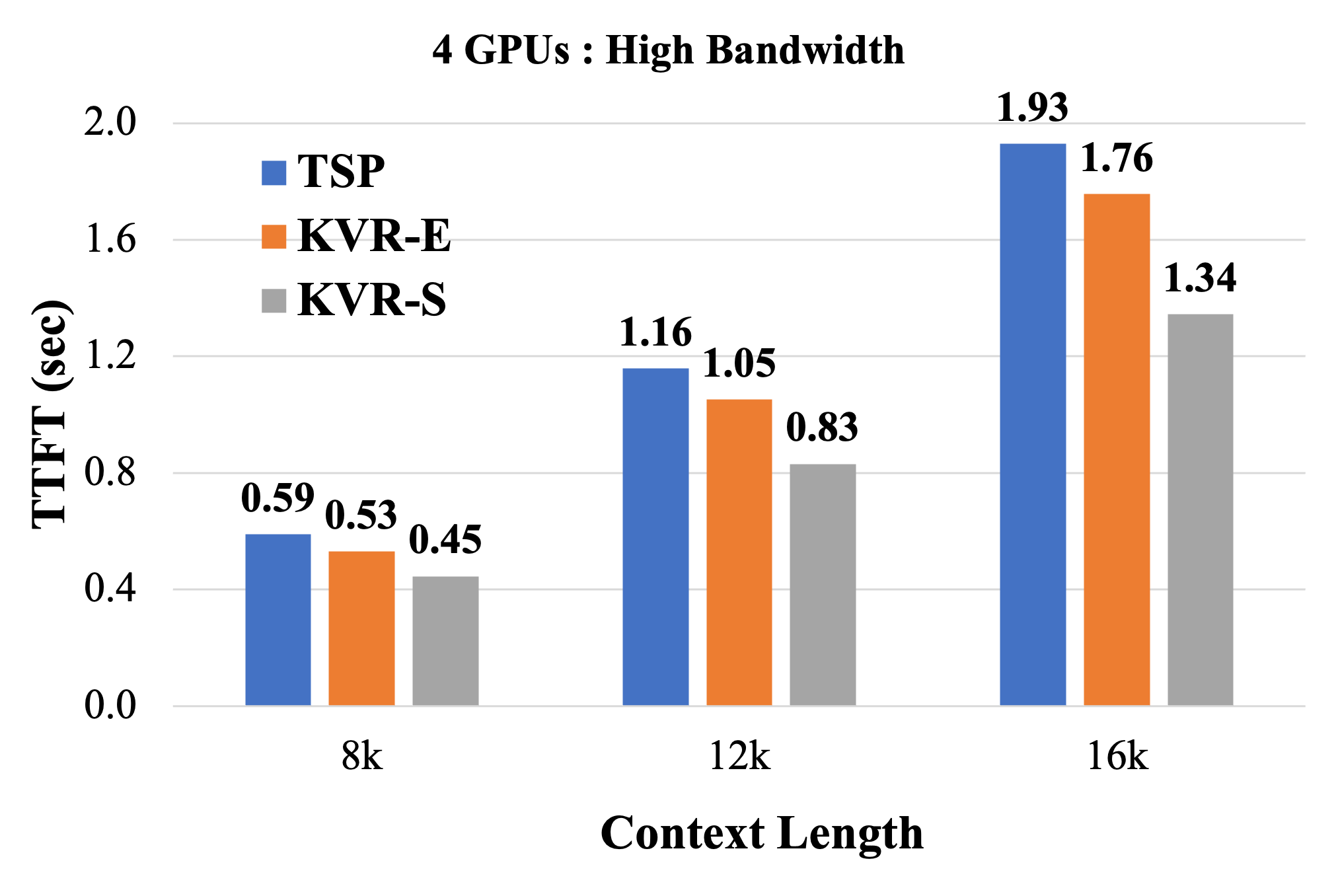}}  
   \hspace{0.05in}
		\subfigure[On a noisy network, \textbf{KVR-S} outperforms \textbf{TSP} even more (41.1\% vs. 45.8\%) than the quiet case in Fig.~\ref{ttft:result_llama7b} (c).]
		{\includegraphics[width=2.2 in]{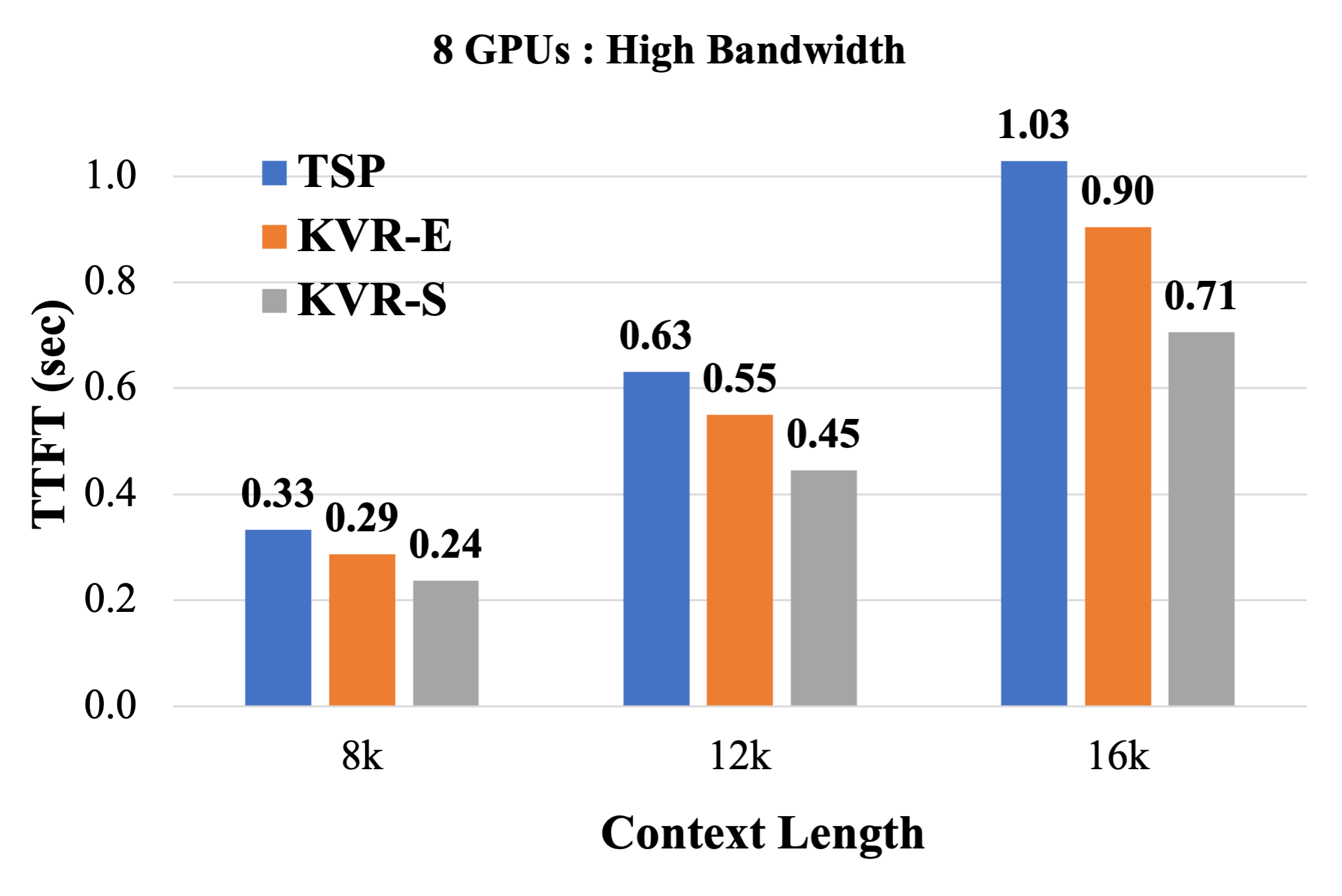}} 
  \hspace{0.05in}
		\subfigure[In terms of the TTFT overhead, \textbf{TSP} is most affected (up to 11.8\%), and  \textbf{KVR-E} is least influenced (up to 2.7\%).]
		{\includegraphics[width=2.2 in]{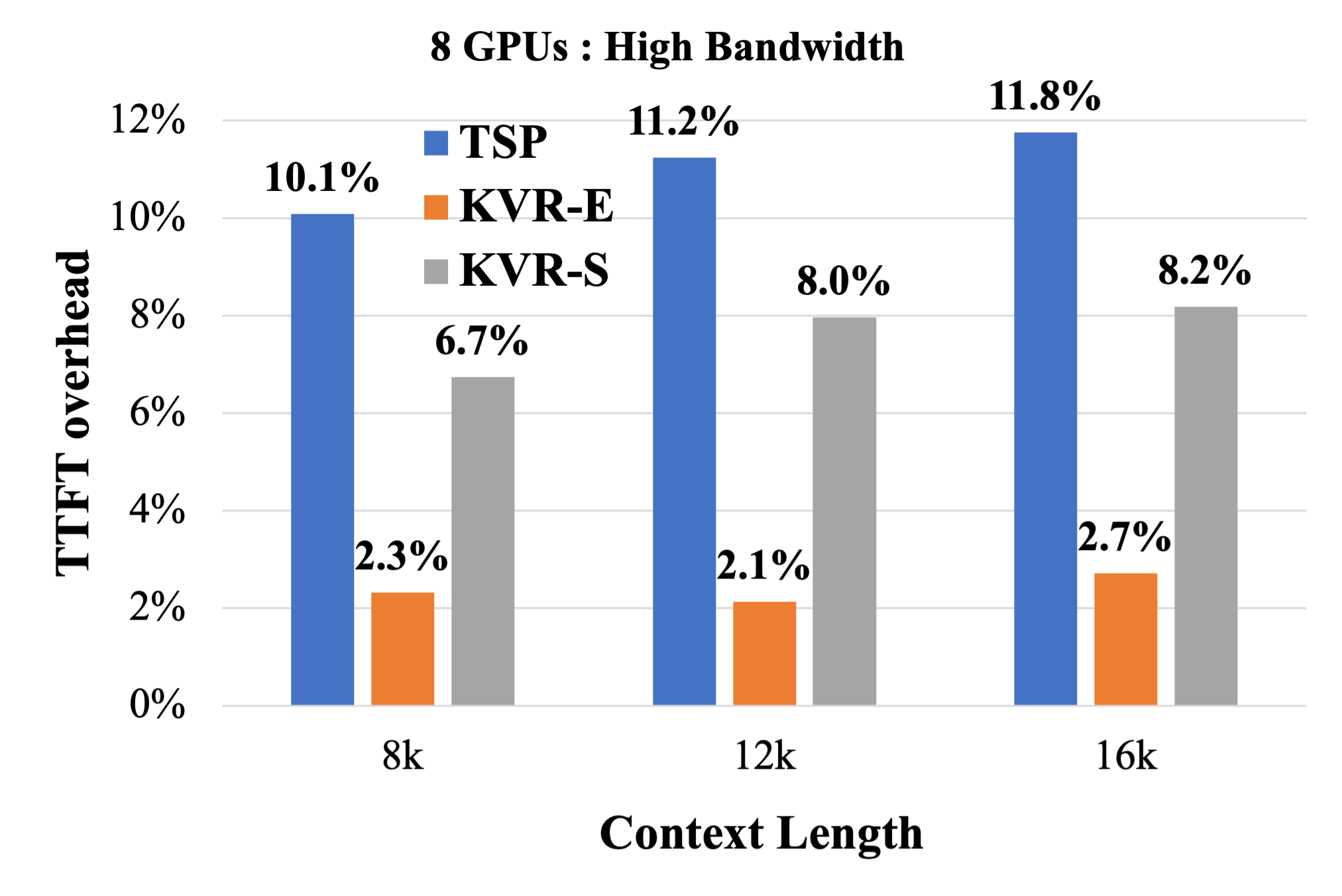}} 
 }  
 
\vspace{-0.1 in}
\caption{
On a noisy high bandwidth network, \textbf{KVR-S} still outperforms  \textbf{TSP} by even wider margin than the case in Figs.~\ref{ttft:result_llama7b} (b, c), and shows high tolerance against the other noisy traffics. Yet, in terms of absolute tolerance,  \textbf{KVR-E} appears to be the best.}
\label{kvr:non_uniform}
    \vspace{-0.1 in}
\end{figure*}

\section{Experimental Results}
\label{kvr::exp}
We used  PyTorch 2.0~\cite{pytorch_eager} and NCCL 2.14 to enable \prjname in Huggingface LLaMA 7B and Falcon 7B~\cite{touvron2023llama,falcon40b}.
All our experiments were done on a single node with $8\times$ NVidia A100 GPUs, and under high (300GB/s) and low (10GB/s) bandwidth setups. Note that we turned off  the high-speed CUDA-direct link~\cite{nccl}  to configure the low bandwidth environments.


We used FP16 for the inference. 
We compared \prjname with Tensor/Sequence Parallelization (\textbf{TSP})~\cite{li-etal-2023-sequence,shoeybi2020megatronlm,patel2023splitwise}. Note that \prjname is applicable to any LLM with causal attention and does not alter any task accuracy. 
For ablation, we created a few variants of \textbf{KVR} as below. 

\begin{table}[!h]
\setlength{\tabcolsep}{4 pt}
\centering
\begin{threeparttable}
\begin{tabular}{c|c } 
\textbf{KVR-E} &  with even context partitioning\\
\textbf{KVR-S} &   with searched context partitioning\\
\textbf{KVR-P} &  with predicted/interpolated context partitioning \\
\end{tabular} 

\end{threeparttable}
\vspace{-0.05 in}

\label{kvr:kvr}
\vspace{-0.1 in}
\end{table}



\textbf{Acceleration:}
Our   results are presented in Figs.~\ref{ttft:result_llama7b} and~\ref{ttft:result_falcon7b} with multiple context lengths and GPU counts. From Figs.~\ref{ttft:result_llama7b} (a-c), we can see \textbf{KVR-S} (even \textbf{KVR-E}) consistently outperforms \textbf{TSP}. And, \textbf{KVR-S} can deliver larger speed up (over 40\%) with longer contexts and more GPUs, and the speedup gain is even higher on low bandwidth (10GB/s) network as in (e, f). Also, note that \textbf{TSP} hit the out-of-memory error for 16k contexts on 2 GPUs, apparently consuming more memory.
Fig.~\ref{ttft:result_falcon7b} shows the similar results with 8k context lengths, but  
speedups are observed only with \textbf{KVR-S} for 4k context.
Fig.~\ref{ttft:result_llama7b} (d) compares the scalabilities of \textbf{TSP, KVR-E}, and \textbf{KVR-S} against two lower bounds: $\textbf{TTFT}(p)$ is the same as \textbf{KVR-S} without any communication (so practical lower bound), and 
$\textbf{TTFT$^*$}(p)$ is from Eq.~(\ref{kvr:attn_p}) (so theoretical lower bound), which leads to the following observations:
\begin{itemize}  
\vspace{-0.1 in}    
   \item $\textbf{TTFT$^*$}(p)$ is very tight to $\textbf{TTFT}(p)$, until the non-parallelizable parts become dominant, like on 8 GPUs.
   \item \textbf{KVR-S} gets much closer than  \textbf{TSP} to  $\textbf{TTFT}(p)$.
   \item  \textbf{KVR-S} is up to 17\% away from $\textbf{TTFT}(p)$ in our tests.
 \vspace{-0.2 in}    
\end{itemize}
More results with other smaller/bigger LLMs and shorter/longer contexts are available in Appendix~\ref{kvr:more_results}.


\textbf{Context-level Partitioning:}
Fig.~\ref{ttft:llama7b_part} (a) discloses the searched context partitioning for the cases in Figs.~\ref{ttft:result_llama7b} (a-c).
In general, we can see the earlier processes need to consume more contexts, and the later ones consume less, which implies that the wait time for the later processes is less of a concern for the configuration. We can use these breakdowns to build a partitioning lookup table, and linearly interpolate the partitionings for 10k and 14k contexts.
For example, we can interpolate from the breakdowns of 8k and 12k to get the predicted partitioning for 10k on 4 GPUs, which results in $[0.350, 0.255, 0.210, 0.185]$ in terms of ratio. And. it can be done similarly for 12k user contexts as well on 4 and 8 GPUs.
According to our results in Figs.~\ref{ttft:llama7b_part} (b, c), even with 4k intervals, \textbf{KVR-P} with predicted partitioning from interpolation is within 1.3\% of the \textbf{KVR-S} cases with the searched partition configurations  and still outperforms \textbf{TSP}.

\textbf{Point-to-point communication:}
To understand the benefit of point-to-point asynchronous communication of \textbf{KVR} over the \textit{all-gather} operation in \textbf{TSP}, we added a noisy sidecar to generate the bidirectional network traffic between a random pair of adjacent GPUs (i.e., simulating dynamically changing non-uniform network bandwidth),  averaged out multiple TTFTs for the 8k, 12k, and 16k context lengths, and then reported the results in Fig~\ref{kvr:non_uniform}. We found that \textbf{KVR} is much more robust against non-uniform bandwidth among processes: while \textbf{TSP} degraded the TTFTs over 10\% on average due to non-uniform effective bandwidth, \textbf{KVR} has up to 3.7\% impact on TTFT, clearly demonstrating the benefits of the communication mechanism in \prjname. Also,  \textbf{KVR-S} is tuned to the quiet environment, but still outperforms \textbf{TSP} thanks to the point-to-point communication.
\section{Conclusion}
In this work, we propose an efficient parallel LLM inference technique, \prjname, to minimize the time-to-first-token. With the proposed techniques, we observed over 60\% speedup in the first token generation over the existing parallelization schemes and  higher robustness against a non-uniform bandwidth environment.

\section{Impact Statements}
 This paper presents work whose goal is to advance the field of Machine Learning. There are many potential societal consequences of our work, none which we feel must be specifically highlighted here.

\bibliography{main}
\bibliographystyle{icml2024}

\clearpage
\newpage
\appendix

\onecolumn
\begin{table*}[!t]
\begin{center}
\begin{tabular}{c|c|ccc|ccccc}
    \hline
    \multirow{3}{*}{Network}&     &  \multicolumn{3}{c|}{4 GPUs} &  \multicolumn{3}{c}{8 GPUs} \\\cline{3-8}
              &        Context  &  \multicolumn{3}{c|}{Method} &  \multicolumn{3}{c}{Method}   \\ \cline{3-8}
              &        Length  &  TSP & KVR-S  & SpeedUp$\times$  &  TSP & KVR-S  & SpeedUp$\times$\\ \hline
    \multirow{6}{*}{Llama 7B} &1k		& 0.107	& 0.097	& 1.11	&	0.117	& 0.098	& 1.19\\ 
                              &2k		& 0.111	& 0.100	& 1.11	&	0.117	& 0.103	& 1.14\\
                              &4k		& 0.20	& 0.17	& 1.17	&	0.12	& 0.11	& 1.14\\
                              &8k		& 0.54	& 0.41	& 1.30	&	0.30	& 0.22	& 1.36\\
                              &12k	& 1.06	& 0.76	& 1.39	&	0.57	& 0.41	& 1.37\\ 
                              &16k	& 1.76	& 1.24	& 1.42	&	0.92	& 0.65	& 1.41\\ 
    \hline 
    \multirow{6}{*}{Llama 13B}&1k		& 0.140	& 0.126	& 1.12	&	0.150	& 0.129	& 1.16\\ 
                              &2k		& 0.143	& 0.131	& 1.09	&	0.153	& 0.131	& 1.17\\
                              &4k		& 0.32	& 0.29	& 1.12	&	0.19	& 0.16	& 1.17\\
                              &8k		& 0.87	& 0.68	& 1.27	&	0.49	& 0.36	& 1.35\\
                              &12k	& 1.71	& 1.25	& 1.36	&	0.91	& 0.66	& 1.37\\ 
                              &16k	& 2.89	& 2.05	& 1.41	&	1.46	& 1.05	& 1.39\\
    \hline                 
    \multirow{2}{*}{Llama 30B}&1k		& 0.21	& 0.20	& 1.08	&	0.23	& 0.19	& 1.19\\ 
                              &2k		& 0.28	& 0.26	& 1.06	&	0.24	& 0.20	& 1.19 \\
    \hline 
    \multirow{4}{*}{Falcon 1B}&1k		& 0.073	& 0.061	& 1.18	&	0.081	& 0.066	& 1.23\\ 
                              &2k		& 0.067	& 0.060	& 1.12	&	0.079	& 0.065	& 1.23\\
                              &4k		& 0.09	& 0.07	& 1.26	&	0.08	& 0.06	& 1.21\\
                              &8k		& 0.25	& 0.19	& 1.28	&	0.15	& 0.10	& 1.58\\ 
    \hline 
    \multirow{4}{*}{Falcon 7B}&1k		& 0.107	& 0.095	& 1.12	&	0.119	& 0.096	& 1.24\\ 
                              &2k		& 0.117	& 0.103	& 1.13	&	0.118	& 0.099	& 1.20\\
                              &4k		& 0.28	& 0.21	& 1.30	&	0.18	& 0.12	& 1.47\\
                              &8k		& 0.78	& 0.54	& 1.46	&	0.47	& 0.29	& 1.63 \\                                                         
\end{tabular}

\end{center}
\caption{Top-1 accuracy with ImageNet1k: \textbf{\prjname} outperforms other schemes with various pruning rates.}
\label{kvr:more_networks}
\end{table*}

\section{Additional Experiments}
\label{kvr:more_results}
In this section, we present additional results with  a wider range of LLMs using both long and short contexts to confirm that \prjname will generalize well across a broader spectrum of LLMs.
We experimented with Falcon 1B, Llama 13B, and Llama 30B, (in addition to Llama 7B and Falcon-7B from Section~\ref{kvr::exp}) and summarize the results in Table~\ref{kvr:more_networks} where we can observe the followings:

\begin{itemize}  
\vspace{-0.1 in}    
   \item \textbf{KVR-S} consistently outperforms \textbf{TSP} for all cases across 4 and 8 GPUs over high bandwidth network.
   \item The speedup from \textbf{KVR-S} is less with shorter inputs (as attention is less bottlenecked). 
\end{itemize}

Also, we tested Llama 7B with Multi-Query-Attention (MQA) and Group-Query-Attention (GQA)~\cite{ainslie2023gqa} over high bandwidth network and report the results in Table~\ref{kvr:mqa_gqa}.  MQA and GQA~\cite{ainslie2023gqa} are techniques to share keys and values among queries so that the attention part can be computationally more efficient. Accordingly, 
$(K,V)$ projection computation costs will be reduced for \textbf{TSP} and\textbf{KVR}, benefiting both.
In detail, \textbf{TSP} has lower communication costs as it has a fewer K and V matrices to \textit{all\_gather}, and
\textbf{KVR} will have lower communication costs with MQA or GQA, as it leads a smaller $(K,V)$ cache.

Compared with Multi-Head-Attention cases from Fig.~\ref{ttft:result_llama7b} (b-c), overall GQA8 and MQA reduce the TTFT universally. For example, the speedup is as large as 1.22x with MQA. \textbf{KVR} demonstrates marginally better speedup gains over \textbf{TSP} with GQA8 and MQA than with MHA. For the example of 8GPU and 16k context, the speedup over \textbf{TSP} was 1.41x with MHA (see Fig.~\ref{ttft:result_llama7b} (c)), but it becomes 1.48x with MQA and 1.46x with GQA.

\begin{table*}[t]
\begin{center}
\begin{tabular}{c|c|ccc|cccccccccc}
    \hline
    \multirow{3}{*}{Network}&     &  \multicolumn{3}{c|}{4 GPUs} &  \multicolumn{3}{c}{8 GPUs} \\\cline{3-8}
              &        Context  &  \multicolumn{3}{c|}{Method} &  \multicolumn{3}{c}{Method}   \\ \cline{3-8}
              &        Length  &  TSP & KVR-S  & SpeedUp$\times$  &  TSP & KVR-S  & SpeedUp$\times$\\ \hline
                              & 1k		& 0.109	& 0.102	& 1.07		& 0.117	& 0.101 	& 1.17\\
                              & 2k		& 0.107	& 0.102	& 1.05		& 0.120	& 0.101 	& 1.18\\
           Llama 7B           & 4k		& 0.18	& 0.14	& 1.23		& 0.12	& 0.10 	& 1.18\\
            MQA               & 8k		& 0.49	& 0.37	& 1.33		& 0.26	& 0.18 	& 1.44\\
                              & 12k		& 0.98	& 0.70	& 1.41		& 0.51	& 0.35 	& 1.45\\
                              & 16k		& 1.65	& 1.16	& 1.43		& 0.84	& 0.57 	& 1.48\\
    \hline  
                              &1k		& 0.112	& 0.102	& 1.10		& 0.119	& 0.101 	& 1.18\\
                              &2k		& 0.113	& 0.102	& 1.12		& 0.118	& 0.103 	& 1.15\\
           Llama 7B           &4k		& 0.18	& 0.15	& 1.20		& 0.12	& 0.11 	& 1.15\\
            GQA8              &8k		& 0.50	& 0.38	& 1.32		& 0.27	& 0.19 	& 1.42\\
                              &12k		& 1.00	& 0.72	& 1.39		& 0.52	& 0.36 	& 1.42\\
                              &16k		& 1.67	& 1.16	& 1.44		& 0.86	& 0.59 	& 1.46\\
 
\end{tabular}

\end{center}
\caption{Top-1 accuracy with ImageNet1k: \textbf{\prjname} outperforms other schemes with various pruning rates.}
\label{kvr:mqa_gqa}
\end{table*}

\begin{table}[!h]
\begin{center}
\begin{tabular}{c|c|cc|cccccc}
    \hline
       Context   & base  & \multicolumn{2}{c|}{10GB/s} &  \multicolumn{2}{c}{1GB/s} \\ \hline
       Length   &  1 GPU & 2 GPU & 4 GPU & 2 GPU & 4 GPU\\ \hline
        1k	& 0.10	&	0.10	& 0.10	& 0.11 &	0.19 \\
        2k	& 0.24	&	\textbf{0.16}	& \textbf{0.19}	& \textbf{0.21} &	0.35 \\
        4k	& 0.65	&	\textbf{0.38}	& \textbf{0.36}	& 0.84 &	0.93 \\
        8k	& 1.95	&	\textbf{0.99}	& \textbf{0.72}	& \textbf{1.31} &	2.06 \\
        12k	& 3.95	&	\textbf{1.82}	& \textbf{1.15}	& \textbf{2.28} &	\textbf{2.30} \\      
\end{tabular}

\end{center}
\caption{Top-1 accuracy with ImageNet1k: \textbf{\prjname} outperforms other schemes with various pruning rates.}
\label{kvr:parallel_inf}
\end{table}

\begin{table}[!t]
\begin{center}
\begin{tabular}{c|c|c|cccccc}
    \hline
    method &  rank/gpu   & partition & size \\ \hline
    \multirow{4}{*}{TSP}  &  0 &	Antibiotics are a&	3 \\
      &  1 &	type of medication	&3 \\
      &  2 &	used to treat	&3 \\
      &  3 &	bacterial infections &	2 \\ \hline
        \multirow{4}{*}{KVR}  &  0 &	Antibiotics are a type of &	5 \\
      &  1 &	medication used to	&3 \\
      &  2 &	treat bacterial	& 2 \\
      &  3 &	infections  &	1 \\
\end{tabular}
\end{center}
\caption{Partitioning examples for Table~\ref{kvr:pseudo_code}.}
\label{kvr:pseudo_partition}
\end{table}

\section{Parallel Inference Benefits}

The benefit of parallel LLM inference depends on the input context size (which determines the parallelization gain) and the network bandwidth (which determines the parallelization cost). To understand when \textbf{KVR} (i.e., parallel LLM inference in general) does help or does not, we experimented with Llama 7B on a low-bandwidth setup (10GB/s) and a poor-bandwidth setup (1GB/s) and report the TTFT for each case in Table~\ref{kvr:parallel_inf}. The bold numbers are when it is beneficial to have multi-GPU inference over single-GPU inference in terms of TTFT. One can observe the followings:
\begin{itemize}  
\vspace{-0.1 in}  
\item  Parallel inference is helpful only when the bandwidth is good enough OR the input context is long enough. For example, the bold numbers which indicate when it is beneficial to have multi-GPU inference over single-GPU inference in terms of TTFT form a lower triangle in the table.
\item Even for parallel inference, when the bandwidth is not high enough, using more GPUs is not always helping. For the example of 2K input and 10GB/s, TTFT is 0.16sec with 2GPUs, but it gets worse into 0.19sec with 4GPUs. Such a degradation is more pronounced with 1GB/s network.
\end{itemize}

All above imply that for a given the infrastructure bandwidth, the optimal system for LLM inference can be determined based on the input size distribution of the target application.
A user request needs to dynamically be assigned to 
a system with the right number of GPUs based on the optimization metric (i.e., cost, latency, utilization, and so on).

\section{Pseudo Code and Example}
\label{kvr:pcode_example}
Table~\ref{kvr:pseudo_code} shows the simplified pseudo code for \prjname integration into an existing transformer implementation, which also contrasts it with \textbf{TSP}. Table~\ref{kvr:pseudo_partition} illustrates one plausible partitioning with \textbf{TSP} and \textbf{KVR} for the example in Table~\ref{kvr:pseudo_code}, underscoring its difference from \textbf{TSP}.

\begin{table}[!t]
\begin{lstlisting}[language=Python]
input = `Antibiotics are a type of medication used to treat bacterial infections`

if method=='tsp':
    context = even_context_partitioning(input, rank, world_size)
elif method=='kvr':
    context = kva_context_partitioning(input, rank, world_size)
else:
    context = input
 
def forward(context, mask, rank, world_size, method, KV_cache=None):

    if method=='kvr' and rank>0:
        KV_cache = net_recv(rank-1)            

    if method=='tsp':
        Q = q_proj(context)
        local_K = k_proj(context)
        local_V = v_proj(context)
        K, V = net_all_gather(local_K, local_V)       
    else: #kvr,base
        Q = q_proj(context)
        K = k_proj(context)
        V = v_proj(context)

    if KV_cache:
        K = cat(KV_cache[0], K)
        V = cat(KV_cache[1], V)   
    
    KV_cache = stack(K, V) 

    if method=='kvr' and rank < world_size-1:
        net_send(KV_cache, rank+1)
    
    attn_weights = softmax (matmul(Q, K.T) + mask)
    attn_output = matmul(attn_weights, V)
    attn_output = o_proj(attn_output)

    return attn_output, KV_cache
\end{lstlisting}
\caption{Simplified Pseudo Code with \prjname Integration}
\label{kvr:pseudo_code}
\end{table}

\section{Lookup Table Generation Overhead}
We analytically derive the cost to precompute a partitioning lookup table (which is a one-time job).
Let us assume that there are $N$ GPUs and a $C$ context with size , and we will pick a stride size at each level such that there are 5 values to check for each  as shown in Figure~\ref{kvr:partition}. Let the time taken for each forward pass to measure TTFT be $T$.

At each level, there are $(N-1)^5$ combinations to evaluate. Once the best combination is picked, we can zoom in and repeat the evaluation for $log_{5-1} C$ levels. 
Therefore, the time taken to precompute the lookup table will be  $T (N-1)^5 log_{5-1} C$ .

For instance, if we assume $T=1$ sec, $N=8$, and $C=16$k    for the case in Fig.~\ref{ttft:result_llama7b} (c), it would take approximately 33 hours for an entry. Moreover, each entry can be searched for in parallel, if more GPUs are available. 
In practice, after a few entries, we can seed the search from the interpolated context partitioning with limited scopes for the expedition.

\end{document}